\documentstyle[aps,prl,floats,epsf,twocolumn]{revtex}
\begin{document}
\newcommand{\dm}       {\Delta m^2}
\newcommand{\sinq}      {\sin^2 2\theta}
\newcommand{\nuebar}      {\bar\nu_{\rm e}}
\draft
\preprint{HEP/123-qed}

\wideabs{
\title{Results from the Palo Verde Neutrino Oscillation Experiment}
\author{F.~Boehm$^3$, J.~Busenitz$^1$, B.~Cook$^3$, G.~Gratta$^4$,
        H.~Henrikson$^3$, J.~Kornis$^1$, D.~Lawrence$^2$,
        K.B.~Lee$^3$, K.~McKinny$^1$, L.~Miller$^4$, 
        V.~Novikov$^3$, A.~Piepke$^{1,3}$, B.~Ritchie$^2$, D.~Tracy$^4$, 
        P.~Vogel$^3$, Y-F.~Wang$^4$, J.~Wolf$^1$}
\address{$^1$ Department of Physics and Astronomy, University of Alabama, Tuscaloosa AL 35487 \\
         $^2$ Department of Physics and Astronomy, Arizona State University, Tempe, AZ 85287 \\
         $^3$ Division of Physics, Mathematics and Astronomy, Caltech, Pasadena CA 91125 \\
         $^4$ Physics Department, Stanford University, Stanford CA 94305}

\date{\today}
\maketitle
\begin{abstract}
The $\nuebar$ flux and spectrum have been measured
at a distance of about 800~m from the reactors
of the Palo Verde Nuclear Generating Station
using a segmented Gd-loaded liquid scintillator detector. 
Correlated positron--neutron events from the reaction 
$\nuebar$p$\rightarrow$e$^+$n
were recorded for a period of 200~d including 55~d with one of the
three reactors off for refueling.
Backgrounds were accounted for by making use of the
reactor-on and reactor-off cycles, and also with a novel 
technique based on the difference between signal and background 
under reversal of the e$^+$ and n portions of the events.
A detailed description of 
the detector calibration, background subtraction, 
and data analysis is presented here.
Results from the experiment show 
no evidence for neutrino oscillations. 
$\nuebar\rightarrow\bar\nu_x$ oscillations were excluded at 90\% CL for 
$\dm>1.12\times10^{-3}$~eV$^2$
for full mixing, and $\sinq>0.21$ for large $\dm$.
These results support the conclusion that the observed
atmospheric neutrino oscillations does not involve $\nu_{\rm e}$.
\end{abstract}
}

\section{Introduction}

Results of a long baseline study of 
$\nuebar$ oscillations at the Palo Verde Nuclear Generating Station
are reported here.
The work was motivated by the observation of an anomalous 
atmospheric neutrino ratio $\nu_\mu/\nu_{\rm e}$
reported in several independent experiments
\cite{Fukuda:1994mc,Becker-Szendy:1992ym,Peterson:1999dc}
that can be interpreted as $\nu_\mu$--$\nu_{\rm e}$
oscillations requiring large mixing.
The mass parameter suggested by this  
anomaly is in the range of 
$10^{-2}<\dm<10^{-3}$ eV$^2$
for two flavor neutrino oscillations.

The quantity $\dm$, defined as the difference 
between the square of the 
masses of the mass eigenstates, and the mixing parameter $\theta$
are related to the transition probability $P$ for 
two-flavor $\nu_a\rightarrow\nu_b$ oscillations
(see, for example, \cite{Boehm:1992nn}) by: 
\begin{equation}
P_{\rm osc}(\nu_a\rightarrow\nu_b)=\sin^22\theta\sin^2
\left(\frac{1.27\Delta m^2L}{E_\nu}\right),
\end{equation}
where $E_\nu$ (MeV) is the neutrino energy, $L$ (m) is the 
source--detector distance, and $\dm$ is measured in eV$^2$.

Exploring $\dm$ down to 10$^{-3}$ eV$^2$ requires that the 
quantity $L/E_\nu$ (m/MeV)
has a value of around 200. 
For reactor neutrinos ($E_\nu\sim$5~MeV), a baseline 
of $L\sim$1~km is adequate. Reactor
experiments are generally well suited to study $\nuebar$ oscillations
at small $\dm$; however, they are restricted to the 
disappearance channel $\nuebar\rightarrow\bar\nu_x$.

Reactor antineutrinos have been used for oscillation studies with ever
increasing $\dm$ sensitivity since 1981\cite{Zacek:1986cu,Declais:1995su}.
All of the experiments are based on the large cross section inverse 
beta decay reaction,
$\nuebar$p$\rightarrow$e$^+$n. 
The correlated signature, a positron followed by a
neutron capture,
allows significant suppression of backgrounds.
As the reactor $\nuebar$ yield and spectra are well known\cite{Zacek:1986cu},
a ``near detector'' is not required. 
It is, however, important to control well the detector 
efficiency and backgrounds.

The mentioned considerations have led to the design of the
Palo Verde and Chooz\cite{Apollonio:1999ae} experiments,
which have similar $\dm$ sensitivities. 
While both experiments 
have pursued their goal of exploring the unknown region of
small $\dm$, recent data from Super-Kamiokande\cite{Fukuda:1998mi} 
favor the $\nu_\mu\rightarrow\nu_x$ oscillation channel
over $\nu_\mu\rightarrow\nu_{\rm e}$. This paper reports 
in greater detail results presented earlier\cite{Boehm:1999gk} and
describes the detector calibration, background subtraction,
and data analysis techniques used to extract results on 
neutrino oscillations.

\section{The Experiment}
\subsection{The detector}
The Palo Verde Nuclear Generating Station in Arizona, the largest
nuclear power plant in the USA, consists of three identical
pressurized water reactors with a total thermal power of 11.63~GW.
The detector is located at a distance of 890~m 
from two of the reactors and
750~m from the third at a shallow underground site.
The 32~meter-water-equivalent overburden
entirely eliminates any hadronic component of cosmic radiation
while reducing the cosmic muon flux to
22~${\rm m}^{-2}{\rm s}^{-1}$.
In order to reduce the ambient $\gamma$-ray flux in the laboratory
all materials in and surrounding the detector were selected for low
activity. The laboratory walls were
built with an aggregate of crushed marble, selected for its low content of
natural radioisotopes. Concentrations of 170, 750, and
560~ppb for $^{40}$K, $^{232}$Th, and $^{238}$U 
were measured in the concrete resulting 
in a tenfold reduction of $\gamma$-ray flux
when compared with locally available aggregate.
A low $^{222}$Rn concentration of about 20~Bq/m$^3$ in the lab air
was maintained with forced ventilation. 
Temperature and humidity were controlled 
to ensure stable detector operation.

\begin{figure}[htb!!!]
\centerline{\epsfxsize=3.25in \epsfbox{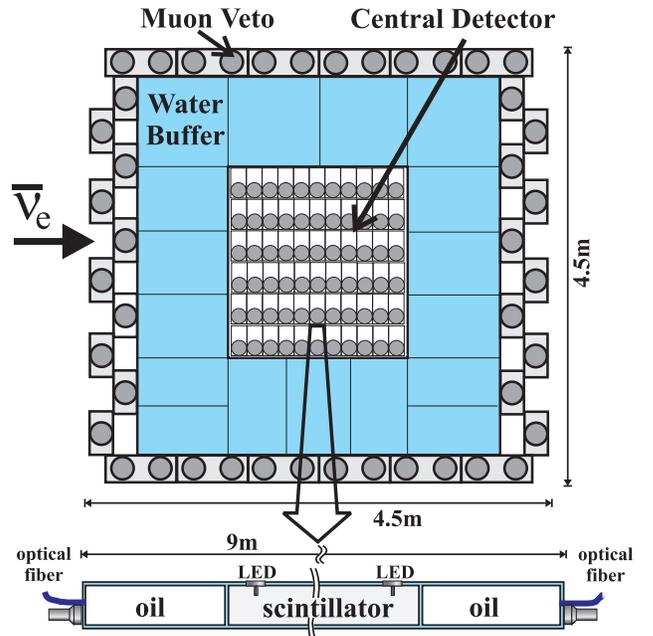}}
\caption{The Palo Verde detector.  One of the 66 target cells 
with PMTs, calibration LEDs, and optical fiber flashers is 
shown lengthwise at the bottom.}
\label{fig:det}
\end{figure}

The segmented detector, shown in Fig.~\ref{fig:det}, 
consists of a 6$\times$11 array of 
acrylic cells dimensioned at 900~cm$\times$12.7~cm$\times$25.4~cm
and filled with a total of
11.34~tons of liquid scintillator.
A 0.8~m long oil buffer at the ends of each cell
shields the central detector from
radioactivity originating in the photomultiplier tubes (PMTs)
and laboratory walls.
The cells were made by cutting and 
bonding large 0.62~cm thick acrylic sheets.
The total acrylic mass in the detector is 3.48~tons.
Each cell is individually wrapped in 0.13~mm thick
Cu foil to ensure light-tightness and is viewed by
two 5-inch low activity PMTs\cite{emi}, one at each end, 
housed in mu-metal boxes.
The target cells are suspended on rollers held in place by thin sheet metal
hangers. All structural materials were dimensioned as lightly as possible
to minimize dead material between cells. Each
cell can be individually removed from the mechanical structure
for maintenance. The detector is oriented such that 
the $\nuebar$ flux is perpendicular to the long axis of the cells.

The liquid scintillator is composed of 36\% pseudocumene, 60\% mineral oil,
and 4\% alcohol, and is loaded with 0.1\% Gd by weight.
This formulation was chosen to yield long light transmission 
length ($11.5\pm0.1$~m at 440~nm),
good stability, high light output, and long term compatibility with acrylic. 
Details of the scintillator development have been published elsewhere 
\cite{Piepke:1999db}.

The central volume is surrounded on the sides
by a 1~m buffer of high purity deionized water (about 105~tons) 
contained in steel tanks 
which, together with the oil buffers at the ends of the cells, serve
to attenuate gamma radiation from the laboratory walls
as well as neutrons produced by cosmic muons passing outside of the detector.
The low {\it Z} of water minimizes the neutron
production by nuclear capture of stopped muons inside the detector
and has a high efficiency for neutron thermalization.

The outermost layer of the detector is an active muon
veto counter, providing 4$\pi$ coverage. 
It consists of 32 twelve meter-long PVC tanks (from the MACRO 
experiment\cite{Ahlen:1993pe})
surrounding the detector longitudinally, and
two endcaps. The endcaps are mounted on a rail system to
allow access to the central detector. The horizontal tanks
are read out by two 5-inch PMTs at each end; the vertical tanks
are equipped with one 8-inch PMT at each end while the endcaps
use 3-inch PMTs. The liquid scintillator used in the veto is a mixture of
2\% pseudocumene and 98\% mineral oil, with a light attenuation
length at 440~nm in excess of 12~m.

\begin{figure}[htb!!!]
\centerline{\epsfxsize=3.25in \epsfbox{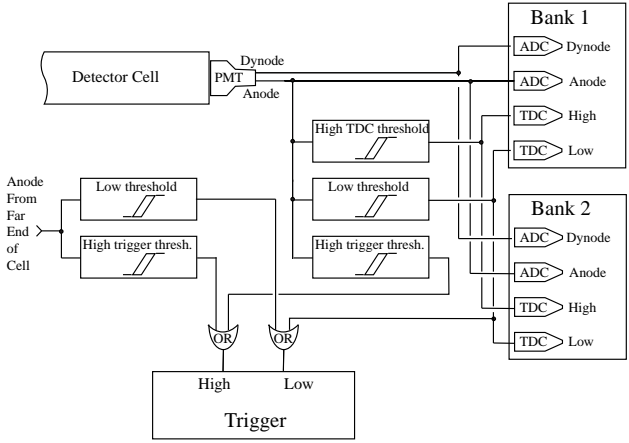}}
\caption{Schematic of the front-end electronics. 
The charge and timing of PMT pulses
are read out by two banks of ADCs and TDCs. The timing information
is discriminated with two thresholds to avoid crosstalk and retain dynamic 
range. Trigger discriminator signals from the two PMTs in each cell are
\textsc{or}'ed and input to the trigger circuit.}
\label{fig:elec}
\end{figure}

A schematic of the central detector's 
front-end electronics is shown in Fig.~\ref{fig:elec}.
Each channel can be digitized by either of two identical banks of 
electronics. The dual bank system allows both 
parts of the sequential inverse beta decay event
to be recorded with no deadtime by switching between banks.
Due to the large dynamic range of energy 
in the data of interest (40 keV to 10 MeV, 
or 1 to 250 photoelectrons typically), 
each PMT has both a dynode and anode output connected 
to ADCs, as well as three discriminator thresholds
for the trigger and TDCs. 
The higher TDC threshold serves to avoid crosstalk from large 
signals in adjacent channels
while the lower threshold allows 
timing information to still be available at the single photoelectron level.
The relative time of arrival 
from each end of a cell is used to reconstruct longitudinal position.
The measured PMT pulse charge at each end, corrected for
light attenuation based on the distance traveled in the cell, 
allows energy reconstruction. 

Each cell is connected to the trigger via the \textsc{or} of 
the discriminated signals from the two PMTs. Signals are
tagged according to two thresholds: a
{\em high} threshold corresponding to 
$\sim$600~keV for energy deposits in the 
middle of the cell and a {\em low} threshold 
corresponding to $\sim$40~keV, or one photoelectron at the PMT. 
The {\em low} trigger threshold also serves as the lower TDC threshold.
The trigger, which has a decision time for each event of around 40~ns,
uses a Field Programmable Gate Array
to search for patterns of energy deposits in the 
central detector,
and can be reprogrammed easily to change trigger 
conditions as needed for calibrations\cite{Gratta:1997cy}.

A veto signal disables the central 
detector trigger for 10~$\mu$s following the 
passage of a muon to avoid most related activity.
Typical veto rates are $\sim$2~kHz.
With each event, the time and hit pattern of the previous muon
in the veto counter is recorded along with 
information as to whether or not the muon passed through the 
target cells. The veto inefficiency was measured to be 
(4$\pm$1)\% for stopping muons (one hit missed) and (0.07$\pm$0.02)\%
for through-going muons (two hits missed).  We note that the small
size of this second quantity with respect to the first is due to 
correlations between incoming and outgoing muons as confirmed by a 
simple Monte Carlo model.

\subsection{The $\nuebar$ signal}
The $\nuebar$ signal is detected via the reaction 
$\nuebar$p$\rightarrow$ne$^+$ as illustrated in Fig.~\ref{fig:sig2} 
further below along with the dominant backgrounds.
Signal events consist of 
a pair of time-correlated subevents: 
(1) the positron kinetic energy ionization and two annihilation 
$\gamma$'s forming the prompt part and 
(2) the subsequent 
capture of the thermalized neutron on Gd forming the delayed part.
By loading the scintillator with 0.1\% Gd, which has a high 
thermal neutron capture cross section,
the neutron capture time 
is reduced to $\sim$27~$\mu$s from 
$\sim$170~$\mu$s for the unloaded scintillator. 
Furthermore, Gd de-excites by releasing an 8 MeV 
$\gamma$ cascade, whose summed energy 
gives a robust event tag
well above natural radioactivity. In contrast, neutron capture on protons
releases only a single 2.2~MeV $\gamma$.

Background is rejected at trigger level using the detector 
segmentation by 
looking for coincidences of energy deposits matching the
pattern of inverse beta decay. 
Each of the subevents of a $\nuebar$ signal
is triggered by scanning the detector for 
a pattern of three simultaneous
hits in any 3$\times$5 subset of the cell
array.  This threefold coincidence, called a {\em triple}, must consist of 
at least one {\em high} trigger hit,
due to either the positron ionization or neutron capture cascade core, 
and at least two additional {\em low}
trigger hits,
resulting from either positron annihilation $\gamma$'s or 
neutron capture shower tails.
The use of identical trigger requirements for the 
two {\em triples} is found to give rise to 
close to an optimal signal to noise ratio.
Five~$\mu$s after finding an initial {\em triple}, 
the trigger begins searching for a delayed {\em triple}. The blank time 
suppresses possible false signals from PMT afterpulsing.
If two {\em triples} are found within 450~$\mu$s of each other, the candidate
$\nuebar$ event is digitized for offline analysis.

\subsection{Expected $\nuebar$ interaction rate}
In order to calculate the expected $\nuebar$ interaction rate
in the detector, the status of the 
three reactors is tracked daily, and the fission rates in the 
cores are calculated based on a simulation code provided by the 
manufacturer of the reactors. This code uses as input the power 
level of the reactors, various parameters measured in the 
primary cooling loop, and the original composition of the core fuel 
elements. 

The output of the core simulation has been checked 
by measuring isotopic abundances
in expended fuel elements in the core;
errors in fuel exposure and isotopic abundances are estimated 
to cause $<0.3$\% uncertainty in the $\nuebar$ flux estimate.
Of the four isotopes  --- 
$^{239}$Pu, $^{241}$Pu, $^{235}$U, and $^{238}$U --- 
whose fissions produce virtually
all of the thermal power as well as neutrinos,
measurements of the neutrino yield per fission and energy spectra 
exist for the first three\cite{Hahn:1989zr,Schreckenbach:1985ep}.
The $^{238}$U yield, which contributes 11\% 
to the final $\nuebar$ rate, is calculated from theory\cite{Vogel:1980bk}.
When the same theoretical method was used to calculate the 
spectrum from the other three isotopes, the theory agreed with experimental
results within 10\%. The contribution of $^{238}$U fission
to the overall uncertainty in $\nuebar$ rate is therefore 
expected to be $\sim$1\%.

\begin{figure}[htb!!!]
\centerline{\epsfxsize=3.5in \epsfbox{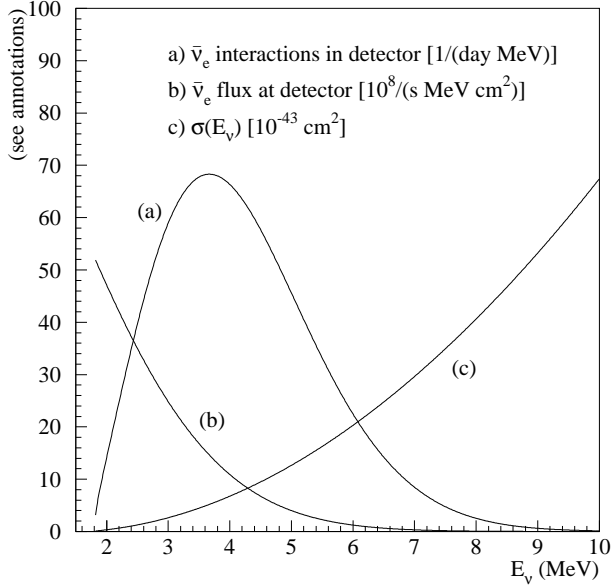}}
\caption{$\nuebar$ flux, inverse beta decay cross section,
and $\nuebar$ interaction spectrum at the detector.}
\label{fig:spec2}
\end{figure}

This calculated $\nuebar$ flux is 
then used to compute the expected rate of $\nuebar$ candidates 
$N_{\nuebar}$ at the detector
as a function of the oscillation parameters $\dm$ and $\sinq$:
\begin{eqnarray}
\nonumber\lefteqn{N_{\nuebar}=
n_{\rm p}\int dE_{\nuebar}\sigma(E_{\nuebar})\eta(E_{\nuebar})\times}    \\
& & \sum_{i=1}^{3}
\frac{{\mathcal I}_{\nuebar,i}(E_{\nuebar})\left(1-P_{{\rm osc},i}(\Delta{\rm m}^2,\sin^22\theta,L_i,E_{\nuebar})\right)}{4\pi L_i^2}
\end{eqnarray}
where $\sigma(E_{\nuebar})$ is the inverse beta decay cross 
section\cite{Vogel:1999zy},
$\eta(E_{\nuebar})$ is the (energy dependent) detector
efficiency, $n_{\rm p}$ is the 
number of target free protons, and
${\mathcal I}_{\nuebar,i}$ is the source strength of reactor $i$ at distance 
$L_i$ with oscillation probability $P_{{\rm osc},i}$.
In Fig.~\ref{fig:spec2} we show the energy spectrum of the $\nuebar$'s
emitted by a reactor, the $\nuebar$ (energy) differential cross section in the 
detector and the actual interaction rate in the detector target 
before detector efficiency corrections, referred to 
here as $R_{\nuebar}$ (obtained by setting $\eta(E_{\nuebar})$=1).
The energy spectrum actually measured in the detector is the
energy of the positron created by the inverse beta decay.
This spectrum is approximately $E_{\nuebar}-1.8$~MeV, slightly 
modified by the kinetic energy carried away by the neutron 
($\sim$50~keV).

Previous short baseline experiments which measured the rate
of $\nuebar$ emission by reactors have found good 
agreement between calculated and observed neutrino flux
by using largely the same method of calculation. 
A high statistics measurement at Bugey\cite{Declais:1995su},
in particular, found excellent agreement both in spectral 
shape ($\chi^2/{\rm n.d.f.}$=9.23/11) 
and in absolute neutrino yield 
(agreement better than 3\%, dominated by systematic errors).
These previous generation experiments prove that the reactor
antineutrino spectrum, i.e. the $\nuebar$ flux at the distance $L$=0, 
is well understood.

\begin{figure}[htb!!!]
\centerline{\epsfxsize=3.5in \epsfbox{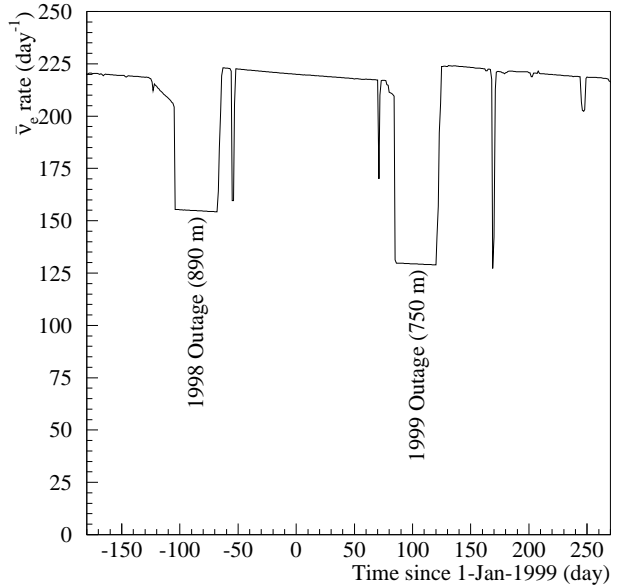}}
\caption{The calculated $\nuebar$ interaction rate in the detector 
target.  The two long periods of reduced flux from reactor refuelings
were used for background subtraction.
The decreasing rate during the full power operation is a result
of the changing core composition as the reactor fuel is burned.}
\label{fig:rates}
\end{figure}

The expected $\nuebar$ 
interaction rate in the whole target, 
both scintillator and the acrylic cells, is plotted
in Fig.~\ref{fig:rates} for the case of no oscillations from 
July 1998 to October 1999. 
Around 220 interactions per day are expected with all three units
at full power.
The periods of sharply reduced rate occurred when one
of the three reactors was off for refueling, 
the more distant reactors each contributing  approximately 
30\% of the rate and the closer reactor the 
remaining 40\%. The short spikes of decreased rate are due to 
short reactor outages, usually less than 
a day. The gradual 
decline in rate between refuelings is caused by fuel burnup,
which changes the fuel composition in the core and the relative 
fission rates of the isotopes, thereby affecting slightly the spectral 
shape of the emitted $\nuebar$ flux.

\section{Calibration}
In order to maintain constant data quality during running,
a program of continuous calibration and monitoring of all 
central detector cells is followed.
Blue LEDs installed inside each cell are used for relative
timing and position calibration. 
Optical fibers at the end of each cell, also illuminated by blue LEDs, 
provide information about PMT linearity
and short term gain changes. LED and fiberoptic scans
are performed once a week.
Radioactive sources are used to map the 
light attenuation in each cell, for absolute energy calibration, and to 
determine detection efficiencies for positrons and neutrons. 
A complete source scan is undertaken every 2--3 months. 

\subsection{LED and optical fiber calibrations}
As seen in Fig.~\ref{fig:det}, 
every cell of the central detector has two LEDs, one at each end at a 
distance of 90 cm from the PMTs. These blue LEDs,
which provide fast light 
pulses with a rise time comparable to scintillation light, 
are used for timing calibrations needed for 
position reconstruction along the 
cell's axis. 

\begin{figure}[htb!!!]
\centerline{\epsfxsize=3.4in \epsfbox{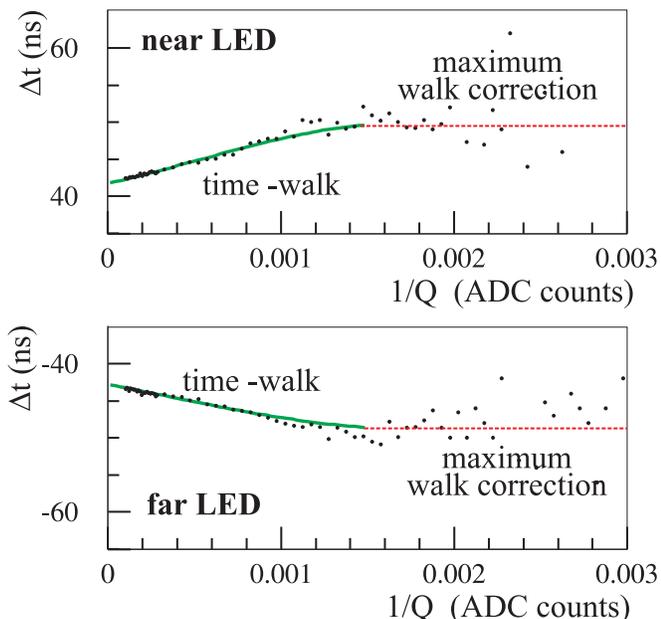}}
\caption{Time difference $\Delta t$ as function of $1/Q$ for
near- and far-end LEDs of a cell, illustrating the {\em time-walk} effect. 
For values of $Q$ close to the discriminator threshold the 
{\em time-walk} correction is kept at a constant (maximum) level.}
\label{fig:dtwalk}
\end{figure}

The difference in 
pulse arrival time between the two PMTs of a cell $\Delta t$ is 
described as a function of the position $z$ with an effective
speed of light $c_{\rm eff}$, an offset $z_0$
and a small nonlinear correction $f(1/Q)$: 
\begin{equation}
\Delta t = (z-z_0)/c_{\rm eff}+f(1/Q_n,1/Q_f).
\end{equation} 
The correction $f(1/Q_n,1/Q_f)$, 
a function of both near and far 
PMT pulse charge $Q$, describes the 
dependence of the pulse height due to {\em time-walk} 
in the leading edge discriminators used in the front-end electronics.  
To extract these calibration parameters and 
compensate for the {\em time-walk} effect, a 
third order polynomial is fit to $\Delta t$ versus $1/Q$ 
(see Fig.~\ref{fig:dtwalk}).
The intercepts at $1/Q = 0$ for the two LED positions provide $c_{\rm eff}$
and $z_0$, while the slopes are used to parameterize the 
{\em time-walk} correction. 

In order to check the suitability of longer wavelength 470~nm LED light 
to measure timing properties 
of $\sim$425~nm scintillation light, 
data taken with a $^{228}$Th source 
at several longitudinal positions
were reconstructed with the LED timing calibration parameters.
Comparing the 
reconstructed positions with the actual
source positions, the effective speed of light measured with the
LED system was found to be on average 
3.6\% lower than that with the sources. A simulation of the light transport 
in a cell with various indices of refraction and attenuation lengths
of the scintillator suggested 
that the small 
discrepancy in $\Delta t$ between LED and scintillation light
was due to the difference in attenuation 
length. 
The correction factor was found to be constant over several months. 
Weekly LED scans are therefore used to correct for short 
term variations in $\Delta t$ and a constant correction 
factor is applied to the effective speed of light.

The fiberoptic system includes 15 blue LEDs, each 
illuminating a bundle of 12 fibers. 
The light output of each LED is measured in two 
independent reference cells with PMTs 
checked to be linear over the whole dynamic range of the LEDs. 
By taking a run which scans through all light intensities and mapping each 
PMT's response relative to the reference cells,
the nonlinear energy response of the PMTs is calibrated.
Low intensities are used to determine the 
single photoelectron gain of 
each PMT, which is used to correct for 
changes from the nominal gain setting of $4\times10^{7}$.

\subsection{Scintillator transparency and energy scale calibration}
In addition to weekly LED and fiberoptic calibrations, 
the energy response of the 
scintillator is measured every three months using a set of sealed 
radioactive sources.
Eighteen 2.4~mm diameter tubes run along the length of the detector allowing 
insertion of the sources adjacent to any cell at any longitudinal position.
The response of each PMT as a function of longitudinal position
is measured by
recording the Compton spectrum from the 2.614~MeV $\gamma$ of a 
$^{228}$Th source at seven different locations along each cell.

\begin{figure}[htb!!!]
\centerline{\epsfxsize=3.5in \epsfbox{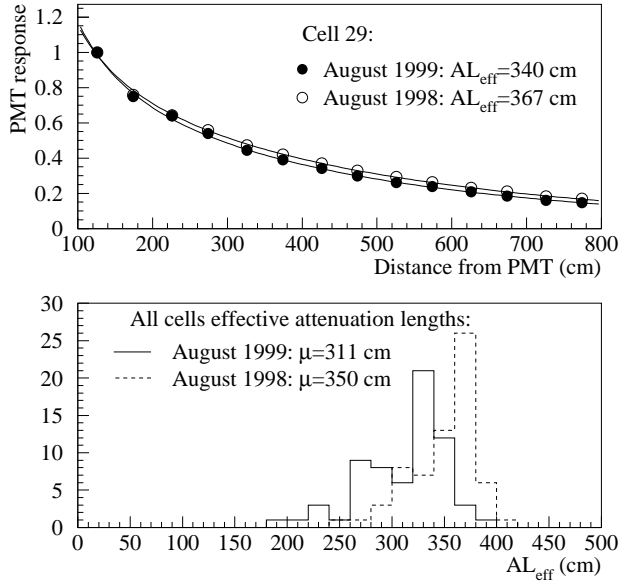}}
\caption{Effect of aging on Gd loaded scintillator.
Top: $^{228}$Th Compton edge position at seven 
different longitudinal locations along a typical cell.  The two curves
are from calibrations taken a year apart.
The curves are normalized at the location nearest to the PMT.
Bottom: Effective attenuation lengths for all 66 cells from 
the two calibrations.}
\label{fig:al}
\end{figure}

Monte Carlo simulation found that 
the half maximum of a Gaussian function 
fitted to the Compton spectrum is relatively 
independent of resolution; this point is therefore used as the 
benchmark of the cell response. 
The response versus distance 
from the PMT, shown in Fig.~\ref{fig:al} for one cell, is then 
fit to the phenomenological function 
${\rm exp}(p_0+p_1z)+{\rm exp}(p_2+p_3z)/z$,
where $z$ is source longitudinal distance from the PMT.
The effective
attenuation length of the scintillator
(including multiple total reflection on the acrylic walls) 
is generally between 3--4 m and over a year 
was found to change on average $\sim$1~mm/day,
demonstrating that the Gd scintillator was remarkably stable.

\begin{figure}[htb!!!]
\centerline{\epsfxsize=3.7in 
\epsfbox{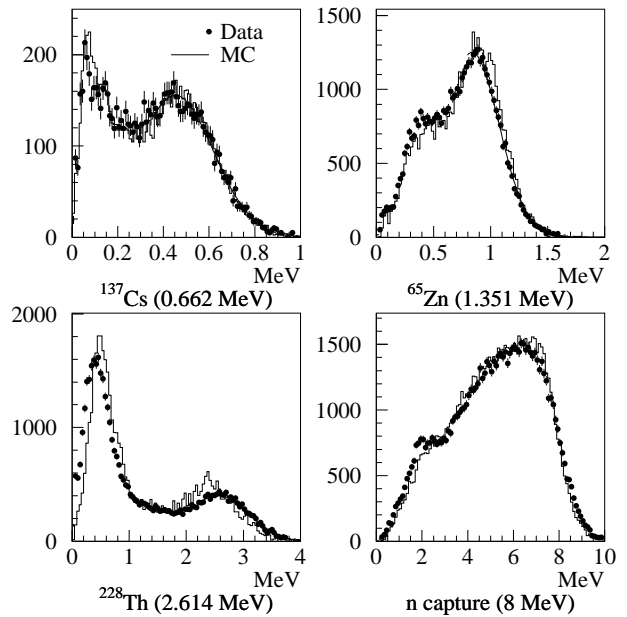}}
\caption{The total energy reconstructed for various sources 
compared for data and Monte Carlo. }
\label{fig:es}
\end{figure}

The overall energy scale was determined 
from the position of the 1.275~MeV peak of a $^{22}$Na source,
and then verified by taking
data with several $\gamma$ sources in different energy ranges: 
$^{137}$Cs (0.662~MeV), $^{65}$Zn (1.351~MeV), $^{228}$Th (2.614~MeV),
and the capture of neutrons (8~MeV) from an 
Am-Be source.
The gamma cascade from neutron capture was modeled 
according to measurements of the 
emitted spectrum\cite{gdcap}. 
In contrast to homogeneous detectors which measure
total absorption energy peaks,
25\% of the detector target mass 
consists of the inert acrylic
of the cell walls, which absorbs some energy.
The Monte Carlo simulation was therefore used to 
find the correct final distributions of energy detected 
from single and multiple scattering of the $\gamma$'s.
The total energy 
reconstructed for data and Monte Carlo for each source
is plotted in Fig.~\ref{fig:es}. 
The data were matched with Monte Carlo 
simulation for the $^{22}$Na spectrum in Fig.~\ref{fig:esna} 
to find the overall energy scale and to 
the spectra in Fig.~\ref{fig:es} to assure that the 
scintillator response is linear over the energies of interest.
The light yield after PMT quantum efficiency
was found to be $\sim$50 photoelectrons per MeV
in the center of the cells.
The agreement for three of the four 
sources in Fig.~\ref{fig:es} is good,
the exception being $^{228}$Th, in which the data has a consistently 
higher Compton scattering peak than Monte Carlo predicts. 
This discrepancy is consistent across all the data taken and therefore
does not affect the scintillator transparency calibration.

\subsection{Monte Carlo simulation}

The $\nuebar$ efficiency of the detector is a 
relatively strong function 
of event location in the detector and, to a lesser extent, 
of time due to scintillator aging.  
A further complication comes from the trigger efficiency 
being a function of threshold (voltage) while only
energy (charge) is measured.
For this reason 
a Monte Carlo model which included a detailed simulation
of the detector response, including the PMT pulse shape, is used for an  
estimate of the overall efficiency for $\nuebar$
detection.  A variety of measurements was performed
to crosscheck that the Monte Carlo accurately models
the detector response.

The physics simulation program is based on \textsc{geant} 3.21\cite{geant}.
This code contains the whole detector geometry and simulated 
the energy, time, and position
of energy deposits in the detector.
Hadronic interactions are
simulated by \textsc{gfluka}\cite{fluka} and the low energy neutron transport 
by \textsc{gcalor}\cite{gcalor}.
Scintillator light quenching, parameterized as a function 
of ionization density,
is included in the simulation\cite{birks}.

\begin{figure}[htb!!!]
\centerline{\epsfxsize=2.5in \epsfbox{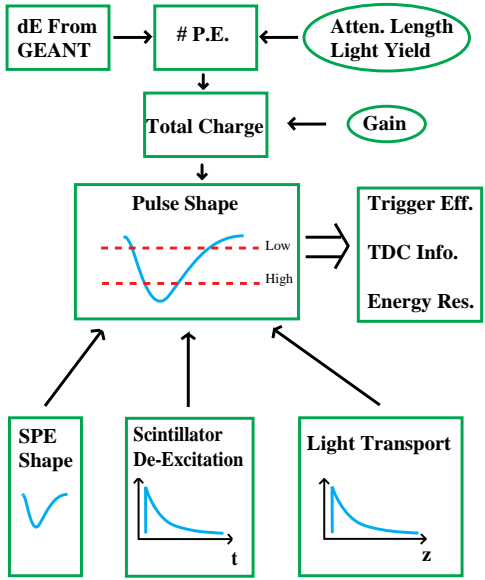}}
\caption{Schematic of the Monte Carlo detector simulation program. 
In order to convert energy deposited into accurate threshold
simulation, calibration information from the detector was 
used to reconstruct PMT pulses.}
\label{fig:mcmodel}
\end{figure}

The event reconstruction program reads the output of this
physics simulation and then applies the 
second step of the Monte Carlo, the simulation of the 
detector response as PMT pulses 
which are then converted into time and amplitude digitizations 
and trigger hits.
A logical scheme of this detailed detector simulation is shown 
in Fig.~\ref{fig:mcmodel}.

The calibrations discussed above 
empirically provide the scintillator light yield
(photoelectron/MeV) and attenuation function for each cell,
which in turn provide the number of photoelectrons 
$\overline{N}_{p.e.}$ expected 
for a simulated energy deposit. The total charge of the pulse then
follows from sampling a Poisson distribution with mean $\overline{N}_{p.e.}$
and folding the number 
of simulated photoelectrons $N_{p.e.}$ with the PMT's nominal
gain, first stage gain variance (10$N_{p.e.}$), and 
cell-to-cell energy scale calibration uncertainty (10\%). 

To simulate the pulse shape, an arrival time is assigned to
each photoelectron, and 
individual photoelectron pulses (whose shape is derived from real data)
are summed
into a final pulse. The 
calculated arrival time of each photon is a 
combination of two processes, scintillator de-excitation and propagation 
along the cell. The latter distribution is parameterized 
by the distance traveled to the PMT, larger distances
giving larger variances, using a light transport simulation 
of $2\times10^7$ photons. The resulting pulse is then analyzed to extract 
TDC and trigger hits.

\begin{figure}[htb!!!]
\centerline{\epsfxsize=3.7in \epsfbox{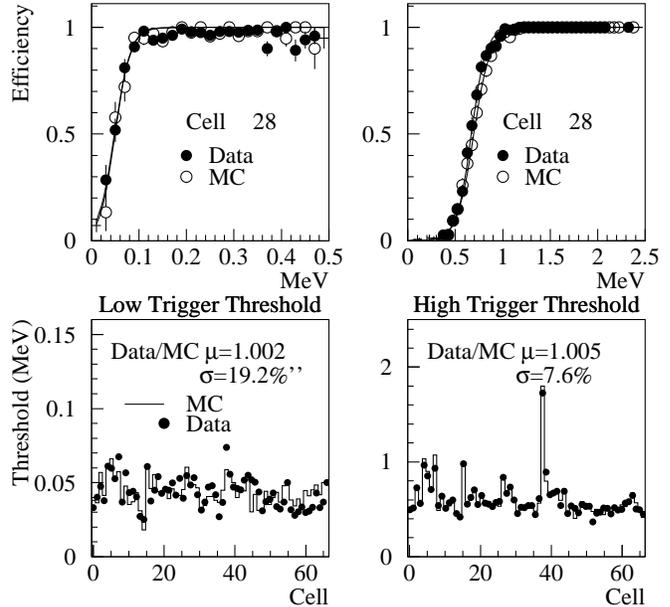}}
\caption{A comparison of the trigger thresholds
from data and Monte Carlo. 
The data were taken with a $^{22}$Na source at the 
center of each cell. 
The top portion shows the efficiency of the 
trigger threshold 
for a typical cell as a function of energy deposited; 
the bottom shows the energy of the 50\% efficiency 
threshold for all 66 cells.}
\label{fig:threshold}
\end{figure}

The Monte Carlo threshold simulation, position reconstruction,
and positron and neutron efficiency predictions were checked
using calibration data. 
The trigger threshold simulation for each cell was compared to data
taken with a $^{22}$Na $\beta^+$ source near the center of each cell.
The trigger conditions were loosened for these data, 
a single {\em low} hit producing a trigger and the event tagged if 
a {\em high} threshold was crossed. 
By plotting the reconstructed energy for each event versus 
the efficiency for a {\em high} trigger tag, an effective {\em high}
trigger threshold in MeV for that location in the cell
was determined. 
The {\em low} threshold was measured similarly. 
The Monte Carlo pulse shape parameters were tuned to 
these data.
A typical cell's trigger threshold efficiency
as a function of energy is shown 
in Fig.~\ref{fig:threshold} for both data and the Monte Carlo.
The trigger threshold, defined as the energy at 50\% efficiency,
is also plotted for all 66 cells.
On average, the thresholds agree to within 1\%. 

\begin{figure}[htb!!!]
\centerline{\epsfxsize=3.5in \epsfbox{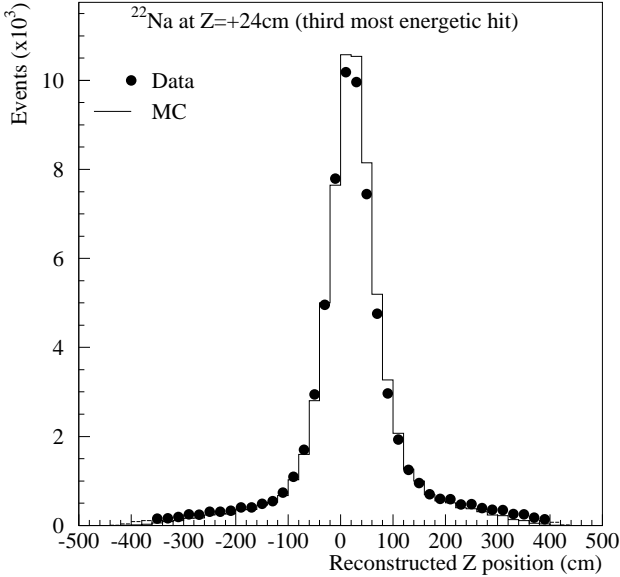}}
\caption{A comparison of the reconstructed $Z$ position of 
an annihilation gamma energy deposit from a $^{22}$Na
run, between data and Monte Carlo. The long tails are due to lower energy 
deposits which do not have TDC 
for position reconstruction, in which case 
relative signal amplitudes are used, resulting in worse 
resolution. Statistical errors in the plot are negligible.}
\label{fig:zpos}
\end{figure}

TDC thresholds were checked by the same algorithm, plotting
the threshold hit efficiency versus reconstructed energy. 
A more direct check of the TDC simulation, however, compares
the position reconstruction for data and 
Monte Carlo simulation. Fig.~\ref{fig:zpos} shows
the longitudinal position 
of the third largest energy deposit in each event 
for a $^{22}$Na calibration run, representing the 
position reconstruction of 
the energy deposited by one of 
the two positron annihilation $\gamma$'s. 
Since these energy deposits tend to be small ($\sim$100~keV), 
some fraction of them have one or both
PMT's responses below the TDC {\em low} threshold. 
These events constitute the tails of the distribution in Fig.~\ref{fig:zpos}
since only the relative signal amplitude was used for position 
reconstruction.
The narrower central peak is populated by events with TDC information 
available. The simulation 
and data agree well, in both resolution and relative frequency of 
the two cases.

\subsection{$\nuebar$ detection efficiency}
The absolute efficiency of the detector for positron annihilations and 
neutron captures was verified using $^{22}$Na and Am-Be 
sources respectively.
The $^{22}$Na source 
emits a 1.275 MeV primary $\gamma$ 
which is accompanied 90\% of the time by a 
low energy positron which annihilates in the source capsule.
The primary $\gamma$ can mimic
the positron ionization of a low energy $\nuebar$ event.
This deposit, in conjunction with the 
positron's 
annihilation $\gamma$'s, closely approximates 
the positron portion of a $\nuebar$ event near the trigger threshold.

\begin{figure}[htb!!!]
\centerline{\epsfxsize=3.7in \epsfbox{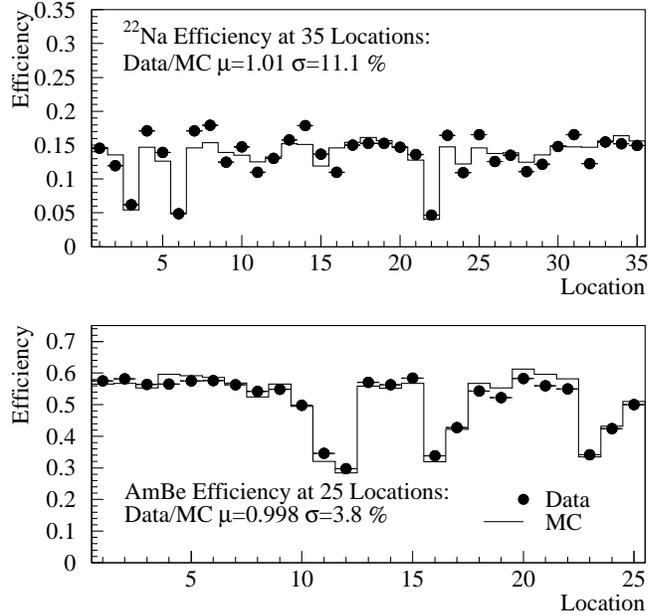}}
\caption{Comparison of data and Monte Carlo 
detection efficiency for Am-Be and $^{22}$Na
source runs at various locations.
Locations at the edges of the detector tend to have lower efficiencies.}
\label{fig:eff}
\end{figure}

In two rounds of data taking,
10 months apart, the $^{22}$Na source was 
inserted into the central detector at 35 locations
chosen to provide a sampling of various distances from the PMTs
and edges of the fiducial volume. 
The source activity is known to 1.5\%, allowing 
determination of an absolute efficiency.
After applying the offline selections used for $\nuebar$ 
prompt {\em triples} and 
correcting for detector DAQ deadtime,
the measured absolute efficiency was 
compared with the Monte Carlo prediction;
the results are summarized in the top portion of Fig.~\ref{fig:eff}.
Good agreement is seen in the average efficiency over all runs, 
and run by run agreement was 11\%. 

\begin{figure}[htb!!!]
\centerline{\epsfxsize=3.7in \epsfbox{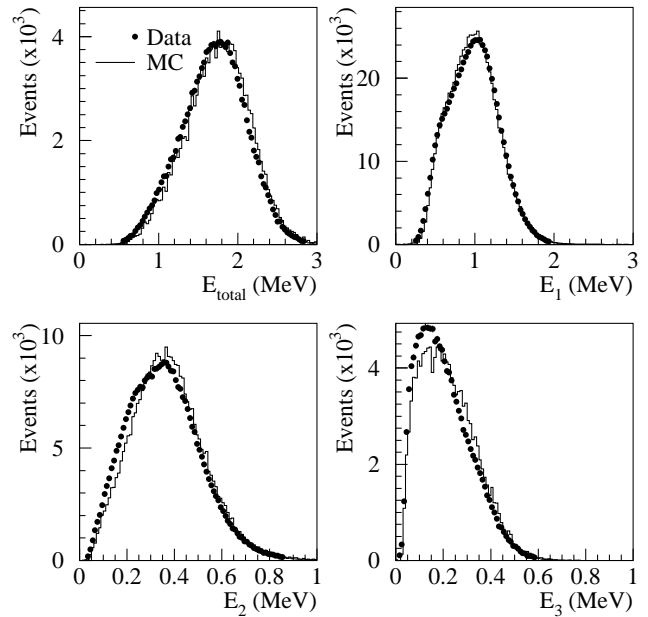}}
\caption{The Monte Carlo simulation and the data
compared for the spectra of total energy and first, second, and third 
most energetic hit ($E_{\rm total}$, $E_1$, $E_2$, and $E_3$)
digitized in the $^{22}$Na calibration runs.}
\label{fig:esna}
\end{figure}

The energy spectra predicted by the simulation 
and measured in the data for the $^{22}$Na runs
were compared. The total 
energy seen in all cells and the energy detected in the 
three most energetic hits are plotted in Fig.~\ref{fig:esna}.
The trigger thresholds can be seen in the spectra: 
the {\em high} trigger threshold is the rising edge  at around 0.5 MeV
in the spectrum of the most energetic hit ({\em E$_1$}), and the {\em low} 
trigger threshold is the rising edge at around 50 keV of 
the third most energetic hit ({\em E$_3$}).

A similar procedure was used to check the neutron capture detection
efficiency. 
The Am-Be neutron source is attached to one end of 
a thin (7.5mm) NaI(Tl)-detector,
which tagged the 4.4 MeV $\gamma$ emitted in coincidence with a neutron. 
The NaI(Tl) tag forces the digitization of the 4.4~MeV 
$\gamma$ as the prompt part of an event and
opens a 450~$\mu$s window for neutron capture;
this is the same coincidence window used in the $\nuebar$ 
runs.

All neutron cuts used for
the $\nuebar$ data selection were applied, and 
the resulting detection efficiency was corrected for detector deadtime 
and a small random coincidence background.
On average, the Monte Carlo efficiency predictions agrees
well over the 25 locations tested
with a run by run agreement of better than 4\%, as shown in the
bottom of Fig.~\ref{fig:eff}.

\begin{figure}[htb!!!]
\centerline{\epsfxsize=3.7in \epsfbox{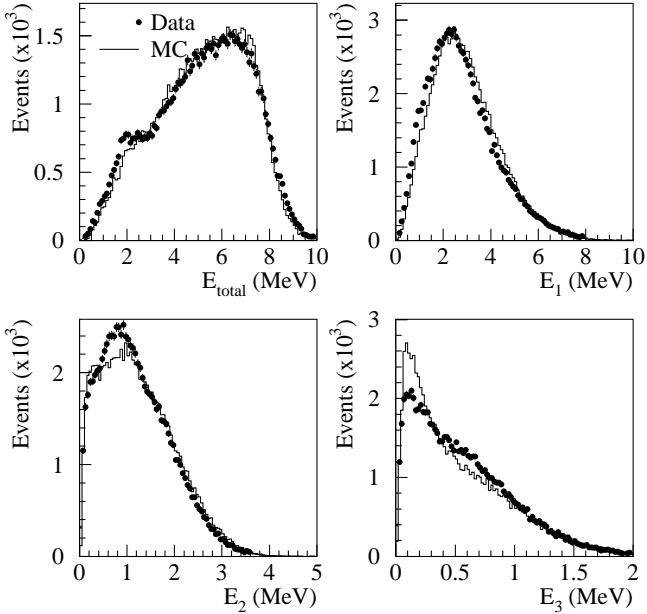}}
\caption{The Monte Carlo simulation and the data
compared for the spectra of total energy and first, second, and third 
most energetic hit ($E_{\rm total}$, $E_1$, $E_2$, and $E_3$) 
digitized in the Am-Be calibration runs.}
\label{fig:esam}
\end{figure}

As with the $^{22}$Na runs, the energy spectra 
predicted by the simulation 
and measured in the data were compared. The total 
energy seen in all cells and the energy detected in the 
three most energetic hits is plotted in Fig.~\ref{fig:esam}.
Note the small peak in $E_{\rm total}$ at $\sim$2~MeV 
arising from neutrons being captured on hydrogen. 
The differences in data versus Monte Carlo 
spectra for $^{22}$Na and Am-Be were taken into account in estimating 
systematic errors.

The Am-Be source emits neutrons with kinetic energies up to 10 MeV,
creating proton recoils in the detector scintillator 
in coincidence with the NaI(Tl) induced trigger.
By digitizing any energy deposits seen during the 
neutron release,
the high ionization density of these recoiling protons 
was used in setting the parameters which control scintillator light quenching
in the simulation.

The above crosschecks verify our ability to accurately generate the
events, model the detector response, reconstruct the events, and 
correctly calculate the livetime of the 
data acquisition (DAQ) system. 
Taken together these procedures complete the task of 
estimating our $\nuebar$ efficiency. 

The Monte Carlo simulation for $\nuebar$ 
events models the expected interactions 
throughout the entire target, including
the acrylic walls of the cells, since there is significant efficiency 
for inverse beta decay originating in the acrylic.
The Monte Carlo simulation yields an average efficiency over the entire 
detector as a function of $\nuebar$ energy. The 
efficiency from the simulation is folded with the 
incident $\nuebar$ spectrum
(which may be distorted by oscillations depending on the 
hypothesis tested), to get the effective efficiency.

\subsection{An independent reconstruction and Monte Carlo}
\begin{figure}[htb!!!]
\centerline{\epsfxsize=3.5in \epsfbox{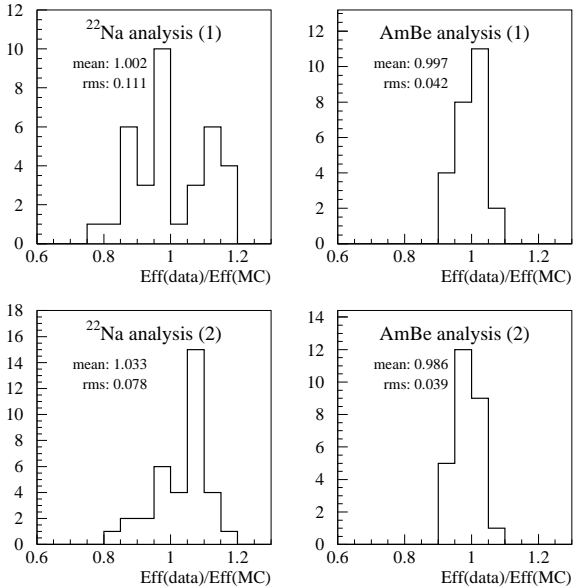}}
\caption{The ratio of predicted to observed efficiency of 
$^{22}$Na and Am-Be for the two reconstruction codes (1) and (2).}
\label{fig:stal}
\end{figure}

A parallel and independent 
event reconstruction and the simulation of the 
detector response has been developed. 
This second version follows 
the same general outline of detector calibration, 
event reconstruction, and simulation described
above, but differs in the algorithms and parameterizations 
used. Major differences include:
\begin{itemize}
\item The functional form for the scintillator
light attenuation is the sum of two exponentials 
${\rm exp}(p_0+p_1z)+{\rm exp}(p_2+p_3z)$,
without $z^{-1}$ in the second term. 
\item The cell response benchmark 
is the 70\% maximum rather than half maximum
of the fitted Compton scattering spectrum.
\item A different parameterization is used for the linearity correction
of the dynode signals.
\item The low threshold parameters are tuned to 
the TDC hit efficiencies rather than 
trigger efficiencies as discussed above.
\item An alternate algorithm for simulating the PMT 
pulse shape was developed and tuned to observed PMT pulse
characteristics.
\end{itemize}
These differences manifest themselves 
as slightly different 
$\nuebar$ efficiency predictions and $\nuebar$ candidate
rates in the data. 
Tests with radioactive sources have been performed to evaluate 
the quality of the second data reconstruction.
The $^{22}$Na and Am-Be efficiency runs shown in 
Fig.~\ref{fig:eff} were reconstructed by the second analysis 
to test its efficiency prediction throughout the detector. 
The ratio of predicted to observed efficiencies over all the 
e$^+$ and neutron runs for the first reconstruction (1) and the second 
reconstruction (2) are plotted in Fig.~\ref{fig:stal}.
While the results presented in the analyses below come from 
the first reconstruction code described above, the 
development of a second simulation and event reconstruction 
offers a useful crosscheck of the systematic uncertainties of the 
results. The differences between the two analyses were used to 
corroborate the estimate of systematic errors.
 
\section{$\nuebar$ selections and backgrounds}
\subsection{$\nuebar$ selection}
The trigger rate for time-correlated events 
(two {\em triples} occurring within 450~$\mu$s) 
is $\sim$1~Hz.
Most of those events are random coincidences of two uncorrelated {\em triple}
hits, which occur individually at 
a rate of $\sim$50~Hz, mostly from natural radioactivity. 
In order to select neutrino events, 
the following offline cuts are applied:
\begin{itemize}
\item The energy reconstructed 
in both prompt and delayed {\em triples}
has at least one hit with $E>$1~MeV and 
at least two additional hits with $E>$30~keV. 
No single hit was allowed to be 
greater than 8~MeV.  
\item The prompt {\em triple} is required to resemble a positron, 
i.e. annihilation $\gamma$'s each less than 600 keV, and together 
less than 1.2 MeV. (This cut is the only one 
which treats the two {\em triples} asymmetrically).
\item At least one 
of the two {\em triples} in the event has more than 
3.5 MeV of reconstructed energy for rejection of
$\gamma$ backgrounds.
\item The prompt and delayed portions 
of the event are correlated 
in space and time (within 3 columns, 
2 rows, one meter longitudinally, and 200 $\mu$s).
\item The event started at least 150~$\mu$s ($\sim$5 neutron 
capture times) after the previous veto tagged muon activity.
\end{itemize}
The trigger and selection efficiencies are summarized in 
the first two columns of Table~\ref{tab:eff}.

\begin{table}[htb!!!]
\begin{tabular}{|l|c|c|c|c|}
 & \multicolumn{2}{c|}{Cumulative} & 
\multicolumn{2}{c|}{Data} \\ 
Cut & \multicolumn{2}{c|}{efficiency} & 
\multicolumn{2}{c|}{Rate (d$^{-1}$)} \\ 
& 1998 & 1999 & 1998 & 1999 \\ \hline
Trigger & 0.271 & 0.328 & 69k & 106k \\ \hline
Selection Cuts & 0.149 & 0.177 & 1k & 1.2k \\
\hline
Live$_{\mu \rm veto}$ & 0.102 & 0.121 & 
\multicolumn{2}{c|}{$<$50, see} \\
Live$_{DAQ}$: &  0.075 & 0.112 & 
\multicolumn{2}{c|}{Table~\ref{tab:swap}} \\
\end{tabular}
\caption{Summary of efficiency and data rates 
after trigger, selections, and live time. The entry ``Live$_{\mu \rm veto}$''
refers to the dead-time induced by the 150~$\mu$s blanking window that accompanies
each muon detected in the veto counter.}
\label{tab:eff}
\end{table}

In addition to corrections for 
selection cut efficiency and trigger efficiency, 
detector livetime is a substantial correction to the 
number of neutrinos seen and deserves some comment. 
Deadtime comes from two sources, the DAQ and the muon veto. 
DAQ livetime is the ratio of the number of {\em triples} 
the DAQ was available for digitizing 
to the total number of {\em triples} the trigger saw.
These numbers are available from trigger scalers. The 
trigger livetime was measured to be $>$99.9\%.
The DAQ live time varies with the {\em triple} rate, and for the 
four data periods was determined to be
73.2\%, 74.4\%, 92.3\%, and 91.8\%
for 1998 full power, 1998 refueling, 1999 full power, and 
1999 refueling, respectively.
The higher livetime in 1999 is due to improvements made in the 
trigger conditions.

The muon deadtime can be further divided into two contributions:
150 $\mu$s of deadtime caused by each muon, which at $R_{\mu}=$ 1990 Hz 
left the detector live 74.2\% of the time; and
muons which interrupted
a neutrino event between the positron and the neutron capture, which 
estimated from the fit parameters of the Monte Carlo capture time
left 92.5\% of events uninterrupted. The total uncertainty in the
calculation of detector deadtime is less than 1\%

\subsection{Backgrounds}
Backgrounds can be separated into two types: correlated and 
uncorrelated. Uncorrelated background
events are due to unrelated {\em triple} hits which 
randomly coincided in the time window allowed. 
Although most of the events collected were random coincidences, almost all 
of this type of background 
is removed by requiring at least one subevent to have
more than 3.5 MeV of reconstructed energy. 
These events do not have a time correlation between prompt and 
delayed subevents ({\em inter-event} time) 
characteristic of neutron capture. 
They have instead a longer time correlation 
determined by the probability that the veto detected no muon
between the prompt and delayed random 
{\em triples}.  At a 2~kHz muon rate, this 
background is seen as a 500 $\mu$s tail under the normal neutron capture 
distribution. By looking at the {\em inter-event} times of the candidate 
$\nuebar$ events at longer time scales, this background 
can be measured. 

\begin{figure}[htb!!!]
\centerline{\epsfxsize=3.7in \epsfbox{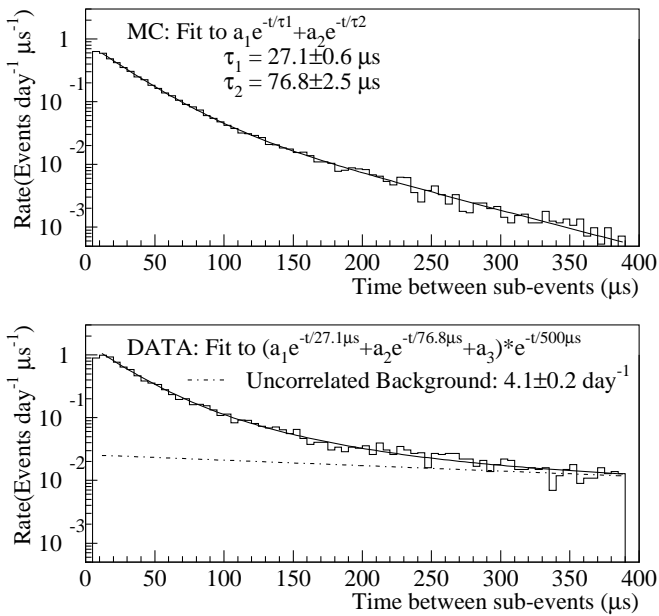}}
\caption{The time elapsed between the prompt and delayed portion of 
neutrino candidate events for Monte Carlo simulation and data.  
The Monte Carlo is fit to two exponentials. The 
data are fit to three exponentials, constrained
to have the time structure of the simulation and 
an additional contribution due to uncorrelated background (dashed
line).}
\label{fig:iet}
\end{figure}

The {\em inter-event} time distribution 
after all neutrino selections (except the 
time correlation cut) is shown in Fig.~\ref{fig:iet}.
The Monte Carlo for a pure neutron capture sample
is empirically fitted to the sum of two exponentials.
There are two time constants due to the 
inhomogeneity of the target: neutrons which remain in the 
scintillator have a 27~$\mu$s capture time, whereas those
which enter the acrylic have a longer capture time due to the 
absence of Gd.
The data {\em inter-event} time distribution is fitted to 
a function of three exponentials with fixed 
time constants consisting of the Monte Carlo fit
$\tau$'s multiplied by a third time constant of 500~$\mu$s.
Integrating the resulting 500~$\mu$s exponential of the uncorrelated 
background in the signal region gives an 
estimate of 4.1$\pm$0.2 events per day, or 9\% of the $\nuebar$ 
candidates being uncorrelated background events.

\begin{figure}[htb!!!]
\centerline{\epsfxsize=3.7in \epsfbox{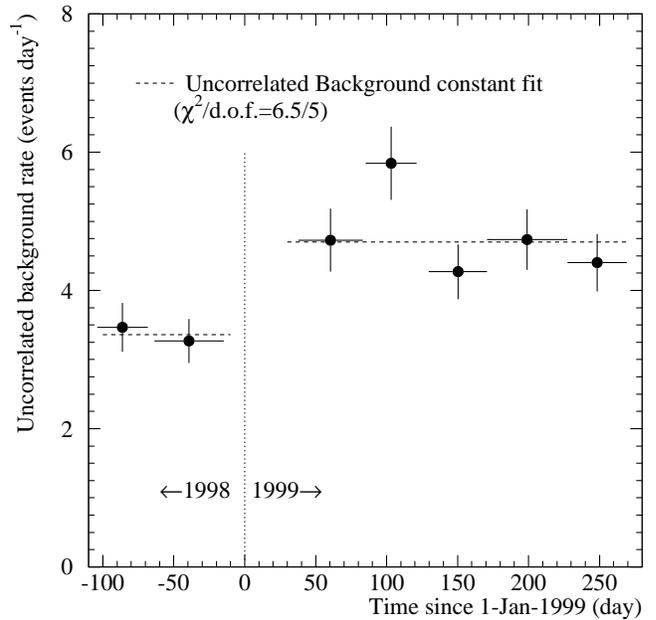}}
\caption{The uncorrelated background rate as 
measured in approximately month-long segments
of the data. Overlain is a fit to a constant within each year's data;
the fit quality is consistent with a stable background.}
\label{fig:ucbg}
\end{figure}

To measure the uncorrelated background in smaller
parts of the data set, the statistical accuracy of the
three exponential fit method 
becomes unacceptably poor. A simpler 
method is therefore used in conjunction with the above fit. 
For {\em inter-event} times longer than 200~$\mu$s,
the $\nuebar$ candidates are dominated by uncorrelated backgrounds.
The integrated number of candidates from 200--400~$\mu$s
is scaled to estimate the number underneath the 
signal region ($<$200~$\mu$s). Using the scaling 
from the fit of the entire data set shown in Fig.~\ref{fig:iet}, 
the uncorrelated background was measured
in approximately month-long intervals as shown in Fig.~\ref{fig:ucbg}.
For both the 1998 and 1999 data sets, the rates are 
found to be stable within statistical errors. 

\begin{figure*}[htb!!!]
\centerline{\epsfxsize=7in \epsfbox{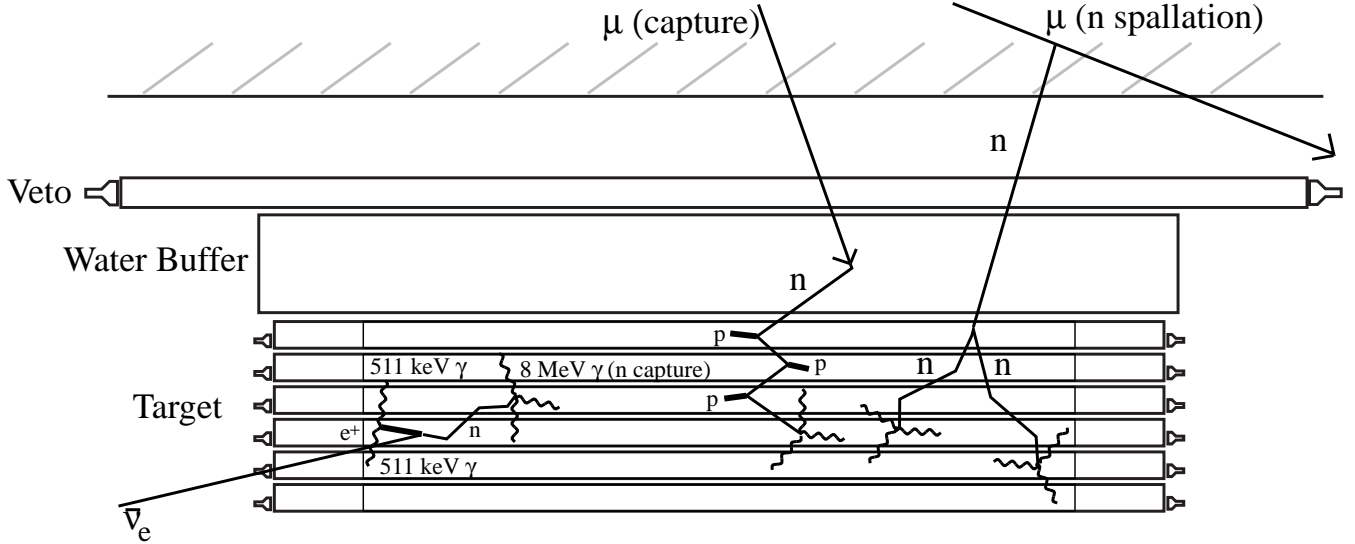}}
\caption{A schematic drawing of the detector showing 
a signal event and two examples of 
interactions which contribute to the correlated background.
Neutrons which cause 
correlated backgrounds come mainly from muons undetected by 
the veto and captured in the water buffer (left), and spallation 
from muons in the lab walls (right). These neutrons in turn 
can either induce more neutrons via spallation (as shown at right) 
or cause three coincident proton
recoils as they thermalize (as shown at left).}
\label{fig:sig2}
\end{figure*}

Correlated backgrounds have the neutron capture {\em inter-event} 
time structure of the $\nuebar$ candidates. These 
events come mainly from 
cosmic muon induced fast neutrons from spallation or muon capture,
as shown schematically in Fig.~\ref{fig:sig2}.
These fast neutrons can either (1) induce more neutrons via spallation, two 
of which can be captured in the detector with one capture mimicking
a positron signature; or (2) they can cause proton
recoil patterns in the central detector which appear as a positron 
signature and then get captured. 
Spallation neutrons originate from muons passing 
through the walls of the lab without hitting the veto detector
or from muons
passing through the detector shielding undetected by the veto.
Muon capture neutrons mainly originate from muons stopping in 
the water buffer without registering in the veto. 

\begin{figure}[htb!!!]
\centerline{\epsfxsize=3.7in \epsfbox{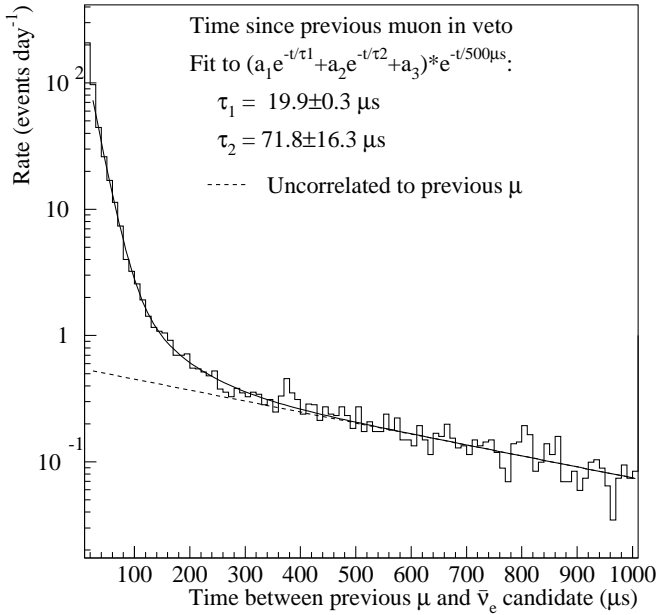}}
\caption{The time between the previous muon traversing the veto counter
and a neutrino candidate event trigger.  
The fit is to a sum of three exponentials.
The dashed portion denotes events uncorrelated to the previous muon,
the rest are correlated background events induced by cosmic muons.
The neutron capture time constants, shorter than the simulation 
prediction for $\nuebar$ events, imply multiple neutron production. }
\label{fig:tslv}
\end{figure}

To illustrate some properties of correlated background, 
Fig.~\ref{fig:tslv} shows the time elapsed since the previous veto hit
for $\nuebar$ candidates, with all selection cuts applied except that 
of the previous muon timing. This distribution is
fit to a three exponential function analogous to that used for the 
{\em inter-event} time fits.
The two time constants for neutron capture are not identical to 
those for $\nuebar$ 
events, but tend to be smaller since after passage of a 
muon there are often more than one neutron in the detector to be 
captured.
The third exponential time constant 
is again constrained to 500~$\mu$s
as expected in a random sampling of events unrelated
to the previous muon. 
Since at very short times
there are other contributions such as muon 
decay, times less than 15~$\mu$s are excluded from the fit.
Muon-induced-neutron backgrounds dominate the candidates in the
first 150 $\mu$s after the previous tagged muons,
motivating the selection cut on $\mu$ timing. 

\begin{figure}[htb!!!]
\centerline{\epsfxsize=3.7in \epsfbox{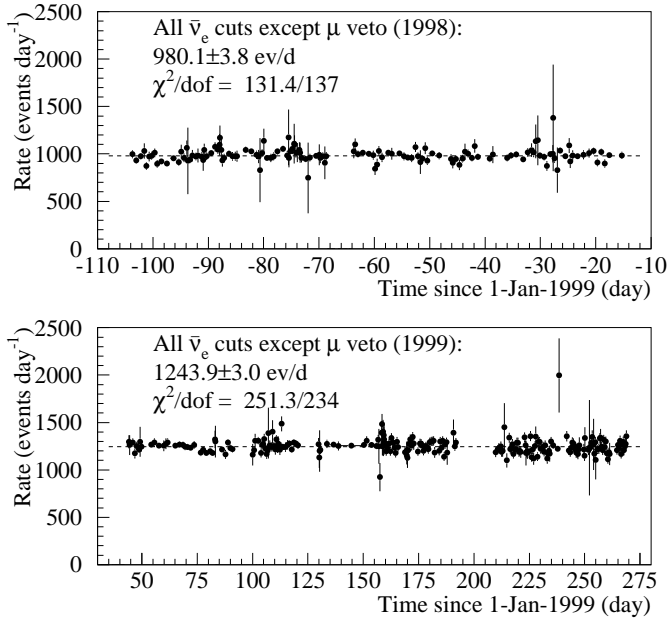}}
\caption{The number of candidates per day in the 373 runs 
used in this analysis, fit to a constant for each year. 
All neutrino selections except the previous muon time cut are used,
leaving a data set dominated by muon-induced-neutron background.
The different rates
in 1998 and 1999 are due to trigger and livetime improvements.}
\label{fig:bg}
\end{figure}

In order to show that 
the correlated background was constant in time,
the previous muon time cut was disabled and 
a plot was made of the $\nuebar$ candidate rate versus time, as shown in
Fig.~\ref{fig:bg}.
When fit to a constant for each year, a $\chi^2$/n.d.f. of 382.7/371
is obtained,
which has a 33\% likelihood, indicating that the detector 
efficiency for correlated backgrounds was stable during 
each year's 
data taking.

Aside from the detector efficiency for background, however, a loss of veto 
efficiency could also cause a fluctuation in background. 
(The rates fit in Fig.~\ref{fig:bg} are with the muon timing
selection disabled, and hence do not vary with veto inefficiency.)
To track veto efficiency,
the veto hit patterns recorded with each event are used.
If a muon hit was recorded only on the bottom of the veto, where
only exiting muons are seen, then the muon must have entered the veto
without recording a hit. By measuring the percentage of these
events a {\em one hit missed} veto inefficiency 
of ($4\pm 1$)\% is found as mentioned above. 
The through-going ({\em two hits missed}) veto inefficiency is measured
to be $0.07\pm0.02\%$ by looking at the rate of $\mu$ tracks
triggered on in the central detector. 
These inefficiencies were
tracked in time to assure their stability.

\subsection{Neutron--$\nuebar$ direction correlation} 
The
neutrons produced in the inverse beta decays will have momenta slightly
biased away from the source, whereas 
no correlation is expected for background. 
This effect is the consequence of momentum conservation which 
requires that the neutron should always be emitted in the forward 
hemisphere with respect to the incoming $\nuebar$. Such a
correlation has been observed already in the G\"{o}sgen 
experiment\cite{goesgen} and again at Chooz\cite{Apollonio:2000jg}.
The theoretical treatment of the effect can be found in 
\cite{Vogel:1999zy}.

\begin{figure}[htb!!!]
\centerline{\epsfxsize=3.7in \epsfbox{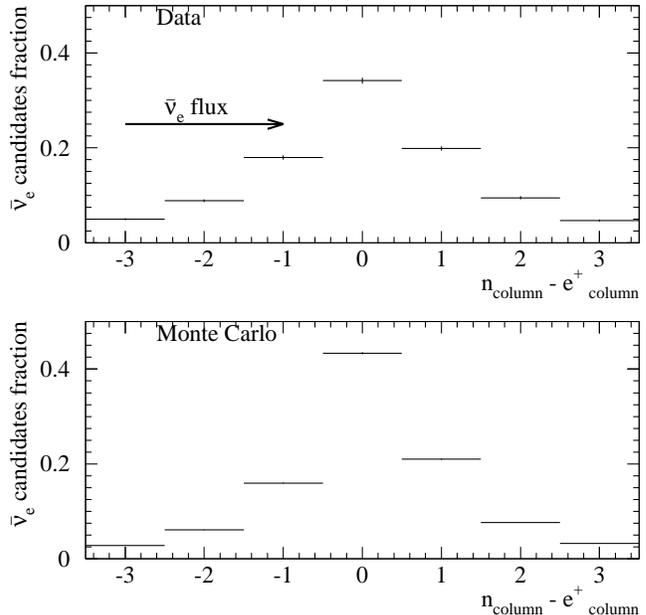}}
\caption{The relative horizontal location
of the neutron capture and positron ionization in the detector target.
The kinematics of the inverse beta decay cause a slight bias in 
the momentum of the neutron from inverse beta decay, seen here 
as an asymmetry in the populations of neutrons
captured one column away or towards the $\nuebar$ source.
Backgrounds should not exhibit such asymmetry.
The Monte Carlo prediction for the pure $\nuebar$ signal, 
normalized to unity, is also plotted.}
\label{fig:asymm}
\end{figure}

The signal to background ratio can be 
independently verified using this effect.
The $\nuebar$ source is to the left of the detector 
in Fig.~\ref{fig:det}. 
The relative horizontal location 
(relative column in the target cell
array) of neutron capture cascade cores versus positron ionizations for 
data and the simulation of the 
$\nuebar$ signal are plotted in Fig.~\ref{fig:asymm}. 
Defining the asymmetry 
$A_{\rm data}=\frac{R-L}{R+L}$ 
in terms of the number of neutrons captured one 
column away from the source $R$ and one column toward the source
$L$, a slight asymmetry $0.050\pm0.017$ is found in the 
data, at 2.9~$\sigma$ significance.
Using the Monte Carlo simulation 
which gives $A_{\rm MC}=0.134$ to 
estimate the portion of the data consisting of $\nuebar$ signal
and assuming the background to be symmetric in this variable,
an effective signal to noise ratio 
\begin{equation}
\frac{S}{N}=\frac{A_{\rm data}}{A_{\rm MC}-A_{\rm data}}
=0.6^{+0.4}_{-0.3}
\end{equation}
is found.
This value agrees well with the ratio of $0.81\pm0.03$ found with
the {\em swap} analysis method described below.

\section{Analysis}

The data set presented here was taken 
from July 1998 to September 1999 in 373 short runs, 
each on average about 12 hours long.
In 1998, 35.97 days of data were taken 
with the three reactors at full power and 31.35 days with one of the 
reactors at a distance of 890~m
off for refueling. The detector was then taken
offline in Jan/Feb of 1999, when DAQ improvements were made
to increase livetime, and the {\em high} trigger thresholds were lowered
by 30\% to increase trigger efficiency. The 1999 data set 
includes 110.95 days with all three reactors at full power and 
23.40 days with the 750 m baseline reactor off for refueling.
Thus, the entire data set has four distinctly different 
periods, with three different baseline combinations and 
neutrino fluxes. 

After all selection cuts there is still substantial background in 
the remaining data set.  The correlated background, coming mainly from 
muon induced neutrons, is difficult to predict and subtract. The yield and
spectrum of neutron spallation is a function of muon flux and energy,
which in turn is a function of depth. While some measurements of 
fast neutron spectra and fluxes have been done in the past, there is 
no model which can consistently predict the fast neutron production.
Below we present three methods used to extract the 
$\nuebar$ signal from data.

\subsection{Analysis with the {\em on-off} method}
The conceptually 
simplest method of subtracting background is to take advantage of 
periods of reduced power levels of the reactor source. 
Ideally all three reactors would be 
down at once allowing for a direct measurement of the background.
However, in practice only one of the three Palo Verde reactors 
was refueled at any given time. 
These reduced power periods occurred twice annually
for about a month. 
Each year's data set is treated independently, subtracting 
1998 {\em off} from 1998 {\em on} and 1999 {\em off} from 1999 {\em on}, since 
the efficiency of the detector
changed between the two years.
By subtracting these data taken at reduced flux 
from the full flux data,
a pure neutrino sample is retrieved albeit containing
the statistical power of only a small portion of the potential data set:
the subtraction is limited by the two months of refueling time and 
treats the $\nuebar$ flux from the two reactors still at full power
as background.

The primary concern arising from use 
of this method, aside from the loss of statistics, 
is guaranteeing that the background rates during the {\em on} and {\em off} 
periods were stable. Both correlated 
and uncorrelated backgrounds were carefully tracked
to ensure stability as discussed above.

\begin{table}[htb!!!]
\begin{tabular}{|l|c|c|}
& 1998 & 1999 \\ \hline
$L$ (m)  & 890 & 750 \\
ON $N_{\rm cand}$ (day$^{-1}$) & $38.2\pm1.0$ & $52.9\pm0.7$ \\
OFF $N_{\rm cand}$ (day$^{-1}$) & $32.2\pm1.0$ & $43.9\pm1.4$ \\
ON-OFF $N_{\rm cand}$ (day$^{-1}$) & $6.0\pm1.4$ & $9.0\pm1.6$ \\
Total efficiency ON (OFF) & 0.0746 (0.0772) & 0.112 (0.111) \\ \hline 
$R_{\rm obs}$ (day$^{-1}$)  & $95\pm19$ & $77\pm14$ \\
$R_{\rm calc}$ (day$^{-1}$) & 63 & 88 \\
\end{tabular}
\caption{Results for the simple {\em on-off} background subtraction
analysis, showing candidate rates in the data $N$ and 
efficiency corrected $\nuebar$ interaction rates 
$R$ observed and calculated.
The data sets for each year are considered independently here.
Uncertainties are statistical only. Systematic errors are estimated 
to be 10\%.}
\label{tab:onoff}
\end{table}

The numerical results of this analysis of the total rate are summarized in 
Table~\ref{tab:onoff}. 
After correcting 
for efficiency (for the no-oscillations 
scenario) and livetime,
the data sets were subtracted to find observed 
neutrino interaction rates in the detector. 
No significant deviation
from the expected neutrino interaction rates 
was found at either baseline distance.

The results from the alternate reconstruction (2) for this analysis 
are shown in Table~\ref{tab:al} for comparison. 
This analysis selects about 5\% more candidates, but also 
gives a correspondingly higher 
efficiency. For this analysis, the uncorrelated background
was measured and removed from the data before the subtraction.

\begin{table}[htb!!!]
\begin{tabular}{|l|c|c|}
& 1998 & 1999 \\ \hline
$L$ (m) & 890  & 750  \\
ON $N_{\rm cand}$ (day$^{-1}$) & $37.3\pm1.2$ & $49.3\pm0.7$ \\
OFF $N_{\rm cand}$ (day$^{-1}$) & $31.6\pm1.2$ & $38.6\pm1.6$ \\
ON-OFF $N_{\rm cand}$ (day$^{-1}$) & $5.7\pm1.7$ & $10.7\pm1.8$ \\
Total efficiency ON (OFF) & 0.0809 (0.0838) & 0.121 (0.121) \\ \hline 
$R_{\rm obs}$ (day$^{-1}$)  & $85\pm20$ & $89\pm15$ \\
$R_{\rm calc}$ (day$^{-1}$) & 63 & 88 \\
\end{tabular}
\caption{Results from the alternate 
reconstruction (2) for the simple {\em on-off} background subtraction
analysis. Uncorrelated background was accounted for in the  
the data before subtraction.
Uncertainties are statistical only. Systematic errors are estimated 
to be 10\%.}
\label{tab:al}
\end{table}

In order to test the results for oscillation 
hypotheses in the two flavor $\Delta m^2-\sin^22\theta$ plane, 
a $\chi^2$ analysis is performed comparing the calculated 
$R_{{\rm calc},ij}$ 
and observed $R_{{\rm obs},ij}$
spectra divided into 1 MeV bins $j$
for each year $i$. 
The spectra used are the prompt energies of the two subtracted data sets. 
At each point in the oscillation parameter plane,
taking into account the changes in detector efficiency 
due to distortions of the neutrino spectrum, the quantity  
\begin{equation}
\chi^2=\sum_{i=1}^{2}\sum_{j}^{E_{\rm bins}}
\frac{\left(\alpha R_{{\rm calc},ij}-R_{{\rm obs},ij}\right)^2}{\sigma_{ij}^2}+\frac{(\alpha-1)^2}{\sigma_{\rm syst}^2}
\label{eq:chisqspec}
\end{equation}
is computed, 
where $\alpha$ accounts for possible global normalization effects due to 
systematic uncertainties (discussed below) 
across both periods and $\sigma_{ij}$ is the 
statistical uncertainty in each bin.
Systematics which can affect spectral shape, mainly energy scale
uncertainty, are negligible relative to the statistical uncertainties
in the analysis.
The function is minimized with respect to $\alpha$.
The point in the physically allowed parameter space with the 
smallest chi-square $\chi^2_{\rm best}$ was found, which represents
the oscillation scenario best fit by the data.

The 90\% confidence level (CL) acceptance region is defined
according to the procedure suggested by 
Feldman and Cousins\cite{Feldman:1998qc} by:
\begin{equation}
\Delta\chi^2=\chi^2(\Delta m^2,\sin^22\theta)-\chi^2_{\rm best}> \\
\Delta\chi^2_{\rm crit}(\Delta m^2,\sin^22\theta)
\label{eq:deltachisq}
\end{equation}
where $\chi^2(\Delta m^2,\sin^22\theta)$ is the minimized 
fit quality at the current point in $\dm-\sinq$ space and
$\Delta\chi^2_{\rm crit}$ is the CL $\chi^2$ cutoff.
Due to the sinusoidal dependence of the expected rates on the 
oscillation parameters and the presence of 
physically allowed boundaries to those parameters,
the cutoff is not simply the 
$\Delta\chi^2$ one would analytically find for 
a three parameter minimization but has to
be calculated for each point in the plane.
To find the $\Delta\chi^2_{\rm crit}$ for a point, the 
experiment is simulated $10^4$ times under the assumption 
that the oscillation hypothesis represented by that point is true. 
For each simulated data set,
a $\chi^2_{\rm best}$ is extracted and a $\Delta\chi^2$  
found for the point. These 10$^4$ $\Delta\chi^2$,
the simulations' fit qualities to the hypothesis,
are then ordered.
The $\Delta\chi^2$ of which 90\%
of the simulations are a better fit is 
a 90\% CL and therefore that oscillation hypothesis'
$\Delta\chi^2_{\rm crit}$. 

\begin{figure}[htb!!!]
\centerline{\epsfxsize=3.7in \epsfbox{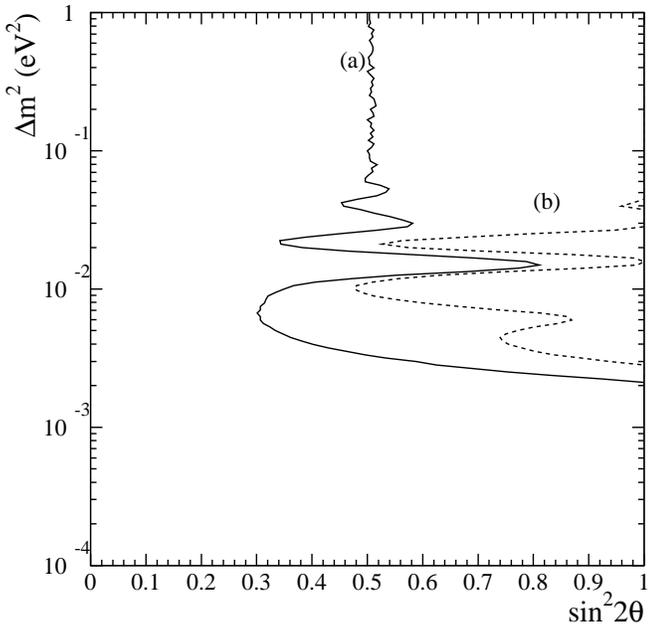}}
\caption{90\% CL limits of the {\em on-off} analysis. 
(a) on-off subtraction 
with normalization constraint, and
(b) using only the spectral shape.}
\label{fig:exclspec}
\end{figure}

The region excluded by the analysis is shown in 
curve (a) of Fig.~\ref{fig:exclspec}.
The results of the $\chi^2$ analysis, including the oscillation 
parameters' best fit to the data, are summarized in 
the first column of Table~\ref{tab:chisq}
further below.
For the {\em on-off} analysis a best fit preferring the 
no-oscillation hypothesis was found.

In addition to the analysis of the absolute $\nuebar$ rates observed,
one can analyze the shape of the spectrum of neutrinos seen independently
of the absolute normalization, thereby relieving the result of most 
systematic uncertainties. 
The $\chi^2$ is calculated 
at each point in the oscillation parameter plane 
as in Eqn.(~\ref{eq:chisqspec}),
with no constraint on normalization ($\sigma_{\rm syst}\rightarrow\infty$).
The same procedure as before is followed in defining 
a 90\% CL region in the $\sinq$--$\dm$ plane.
At large $\dm$ where $\nuebar$ of all energies  
are oscillating many times within the
baseline, the energy spectrum of the incident flux is 
affected only in magnitude. As a result,  
the region excluded in the plane 
does not extend to large $\dm$, as shown in Fig.~\ref{fig:exclspec} (b).

For the spectrum analysis, when the normalization $\alpha$ is left free,
the minimum $\chi^2$ is obtained for $\alpha\sim 2$ (see Table~\ref{tab:chisq}) 
and maximum mixing.   This is clearly an unphysical result since
such large value of $\alpha$ can be excluded to a very high degree of
confidence by the independent efficiency calibrations of the detector
discussed in previous sections.    In addition this result has no effect
on the exclusion plot in Fig.~\ref{fig:exclspec} because, as shown in
Table~\ref{tab:chisq} the no-oscillation hypothesis has actually better
$\chi^2/{\rm n.d.f.}$ than the minimum.   Also the exclusion plots based on 
Eqn.~(\ref{eq:deltachisq}) and either $\chi^2_{\rm best}$ or $\chi^2_{\rm no. osc.}$
are found to be virtually identical.   Furthermore, changing the bin size from 
1~MeV to 0.5~MeV does not appreciably change the exclusion plot, either.

\begin{figure}[htb!!!]
\centerline{\epsfxsize=3.25in \epsfbox{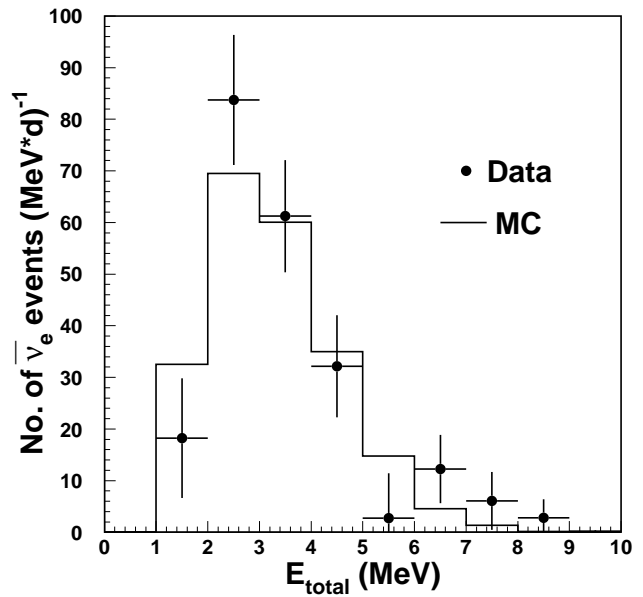}}
\caption{Total energy of the prompt subevent 
($\sim E_{\nuebar}-1.8$ MeV) in 1 MeV wide energy bins
(observed and expected for no oscillation) from the 
{\em on-off} analysis, summed over 1998 and 1999.
The spectra are in agreement with each other, 
with a $\chi^2$/n.d.f. = 8.5/8.
Uncertainties are statistical only.}
\label{fig:onoff}
\end{figure}

Since the analyses reported above and in the following sections finds no 
evidence for neutrino oscillations, the spectra of the two years are added 
and the summed spectrum is plotted in Fig.~\ref{fig:onoff} 
along with the Monte Carlo expectation. 

\subsection{Analysis with the reactor power method}
Part of the statistical limitations of the direct subtraction of the 
preceding analysis is a result of the separation of the data set
by year. By correcting the four periods for efficiency and then 
subtracting the respective reduced flux from full flux periods, 
the subtraction is forced to treat the $\nuebar$ flux from two of the 
reactors as background. A second $\chi^2$ analysis was performed
which effectively uses the full $\nuebar$ flux of the refueling
periods. 

To use the 1998 and 1999 data sets together, 
the change of both signal and background efficiencies are
accounted for.
The $\nuebar$ efficiency difference is found through the detector Monte
Carlo simulation. The efficiency change for background from 1998 to 1999,
which is not necessarily the same as for $\nuebar$, is extracted via the 
high statistics fits of correlated background shown in Fig.~\ref{fig:bg}.
The 27\% increase in background efficiency observed roughly corresponds to
what the Monte Carlo simulation predicts for a background composed 
mainly of double neutron captures. 
Uncorrelated background accounts for less than 10\% of the candidate set;
this background's efficiency changed by a similar amount within
the measurement statistics seen in Fig.~\ref{fig:ucbg}.

The combined data sets are analyzed for oscillation hypotheses by 
calculating the $\chi^2$ summed over runs $i$ (runs with 
less than ten candidates are combined with adjacent runs): 
\begin{equation}
\chi^2=\sum_{i} \frac{\left((\alpha N_{{\rm calc},i} + b)-N_{{\rm cand},i}\right)^2}{\sigma_i^2}+\frac{(\alpha-1)^2}{\sigma_{\rm syst}^2}.
\label{eq:chisqand}
\end{equation}
where $N_{{\rm cand},i}$ is the 
total $\nuebar$ candidate rate,
$N_{{\rm calc},i}$ is the calculated 
rate, $\alpha$ is the overall 
normalization as before, and $b$ is the background rate. 
The background, $b$, is scaled as appropriate for the year but
is otherwise assumed to be constant. The function is minimized at each 
point with respect to $b$ and $\alpha$.
We found no evidence for oscillations and 
the 90\% CL plot, shown in Fig.~\ref{fig:exclyf}, curve (a),
is then constructed around the $\chi^2_{\rm best}$ as
before by comparing $\Delta\chi^2$ with $\Delta\chi^2_{\rm crit}$ 
at each point.
The predictions of signal and background 
from this fit for the no-oscillation hypothesis are 
shown in Table~\ref{tab:and}.
The no-oscillation likelihood and best fit results 
of the $\chi^2$ analysis with this method are summarized
in the third column of Table~\ref{tab:chisq}.

\begin{table}
\begin{tabular}{|l|c|c|c|c|}
Events                    & 1998 ON      & 1998 OFF     & 1999 ON      & 1999 OFF \\ 
(day$^{-1}$)              &              & (890~m)      &              & (750~m) \\ \hline
$N_{{\rm cand}}$	  & 38.2$\pm$1.0 & 32.2$\pm$1.0 & 52.9$\pm$0.7 & 43.9$\pm$1.4 \\
$b$ & \multicolumn{2}{c|}{$19.5\pm1.7$}  & \multicolumn{2}{c|}{$26.3\pm 2.2$} \\ 
$N_{{\rm detected}}$      & 18.7$\pm$2.0 & 12.7$\pm$2.0 & 26.6$\pm$2.3 & 17.6$\pm$2.6 \\ 
\hline
$R_{{\rm obs}}$           & 225$\pm$24   & 140$\pm$22   & 216$\pm$19   & 140$\pm$21 \\ 
$R_{{\rm calc}}$          & 218          & 155          & 218          & 130 \\ 
\end{tabular}
\caption{The rate of candidates $N_{{\rm cand}}$, 
signal $N_{{\rm detected}}$, and background $b$
found by the reactor power analysis,
using a $\chi^2$ analysis of the data set run by run.
The efficiency-corrected total interaction rates $R$ are also listed.
Errors are statistical only. The systematic uncertainty is 
estimated to be 10\%.}
\label{tab:and}
\end{table}

\subsection{Analysis with the {\em swap} method}

A third analysis is used which has the potential of using the 
full statistical power of the neutrino data set by subtracting 
background directly. The method, discussed in more detail 
elsewhere\cite{yfwang}, takes advantage of the asymmetry of 
the prompt (positron) and delayed (neutron capture) subevents 
of the neutrino signal. The 
data selection and trigger
treat the two portions of the event identically with the exception 
of two cuts designed to isolate events with annihilation-like $\gamma$'s
in the prompt {\em triple}. 

The candidates remaining after the selection cuts
can be written as:
\begin{equation}
N = B_{\rm unc} + B_{\rm nn} + B_{\rm pn} 
+ S_{\nu}
\label{eq:n}
\end{equation}
where $B_{\rm unc}$, $B_{\rm nn}$, and $B_{\rm pn}$ are 
uncorrelated, two-neutron, and proton-recoil--neutron-capture backgrounds
respectively; and $S_{\nu}$ is the neutrino signal.
Applying the same neutrino cuts with the positron cuts 
reversed, or {\em swapped},
(such that the positron cuts are now applied to the delayed {\em triple})
gives:
\begin{equation}
N^{\prime} = B_{\rm unc} + B_{\rm nn} + \epsilon_1 B_{\rm pn} 
+ \epsilon_2 S_{\nu}
\label{eq:nprime}
\end{equation}
Since the uncorrelated background and two neutron capture backgrounds
are symmetric under exchange of the prompt and delayed {\em triples}, their
efficiencies with the reversed cuts remain the same.  
The parameters 
$\epsilon_1$ and $\epsilon_2$ denote the relative efficiency 
change for proton recoils and neutrino signal under the swap, 
respectively.

The positron cuts are highly efficient for 
positron annihilation events but have poor efficiency for neutron captures. 
The Monte Carlo simulation is used to estimate $\epsilon_2=0.159$.
Subtracting (\ref{eq:n}) from (\ref{eq:nprime}) leaves the majority of 
the neutrino candidates and only proton recoil background:
\begin{equation}
N - N^{\prime} = (1-\epsilon_1)B_{\rm pn} + (1-\epsilon_2)S_{\nu}.
\end{equation}

To estimate $(1-\epsilon_1)B_{\rm pn}$, it is noted
that the proton recoil spectrum extends beyond
10 MeV, well above the positron energies of the neutrino
signal and other sources of background. 
These measured high energy events can be used to normalize the 
$B_{\rm pn}$ background in the signal using the Monte Carlo ratio:
\begin{equation}
r\equiv\frac{B_{\rm pn}^{\rm MC}({\rm E}_{\rm 1,e^+}<8 {\rm MeV})}
{B_{\rm pn}^{\rm MC}({\rm E}_{\rm 1,e^+}>10 {\rm MeV})},
\end{equation}
where $B_{\rm pn}^{\rm MC}({\rm E}_{\rm 1,e^+}<8 {\rm MeV})$ is the 
fraction of simulated $B_{\rm pn}$ events passing the normal $\nuebar$
selections, and $B_{\rm pn}^{\rm MC}({\rm E}_{\rm 1,e^+}>10 {\rm MeV})$
are the fraction of simulated events in the high energy background region.
Multiplying the ratio $r$ by the measured high energy 
proton recoil rate gives the  $B_{\rm pn}$ background contribution:
\begin{equation}
B_{\rm pn}=rB_{\rm pn}^{\rm data}({\rm E}_{\rm 1,e^+}>10 {\rm MeV}).
\end{equation}

\begin{table*}[t!!!!!!!!]
\begin{tabular}{|l|c|c|c|c|}
Period  & 1998 ON & 1998 OFF &
                1999 ON & 1999 OFF \\
& & 890 m reactor off & & 750 m reactor off \\
\hline
time (days)                & 35.97 & 31.35  & 110.95 & 23.40   \\
$\bar\nu_{\rm e}$ overall efficiency (\%)  &  7.46 & 7.72 & 11.2 & 11.1 \\
\hline $B_{\rm pn}^{\rm data}({\rm E}_{\rm 1,e^+}>10 {\rm MeV})$(day$^{-1}$) & 8.79     & 9.09  & 13.52  & 13.29 \\
$(1-\epsilon_1)B_{\rm pn}$(day$^{-1}$) $\mu$ spallation  & -0.88     & -0.91  & -1.35  & -1.33 \\
$(1-\epsilon_1)B_{\rm pn}$ (day$^{-1}$) $\mu$ capture & 0.58     & 0.58  & 0.86   & 0.86 \\ \hline
$N$ (day$^{-1}$)   & $38.2\pm 1.0$ & $32.2\pm 1.0$ & $52.9\pm 0.7$ & $43.9 \pm 1.4$ \\
$N'$ (day$^{-1}$)   & $24.6\pm 0.8$ & $21.2\pm 0.8$ & $32.3\pm 0.5$ & $31.7 \pm 1.2$ \\ \hline
$\mathrm N_{\nu}$ (day$^{-1}$)& $16.5\pm 1.4$ & $13.5\pm 1.4$ & $25.1\pm 0.9$ & $15.0\pm 1.9$ \\
Total background $B_{\rm unc}+B_{\rm nn}+B_{\rm pn}$ (day$^{-1}$)
& $21.7\pm 1.0$ & $18.7\pm 1.0$ & $27.8\pm 0.6$ & $28.8\pm 1.3$ \\
\hline
$R_{{\rm obs}}$ (day$^{-1}$) & $221\pm 19$ & $174\pm 17$ & $225\pm 8$ & $137\pm 17$ \\
$R_{{\rm calc}}$ (day$^{-1}$)       &   218  & 155   & 218 & 130  \\
\end{tabular}
\caption{Results for the {\em swap}
analysis, including the various background estimates.
Uncertainties are statistical only.}
\label{tab:swap}
\end{table*}

The neutrons which cause the proton recoil background 
are created either by muon capture or spallation in the 
laboratory walls,
or by muons entering the veto counter undetected.
The spectrum of the fast neutrons from spallation
is not well understood. However, such
spectrum can be decoupled somewhat from the resulting
proton recoil spectrum. The expected backgrounds were simulated 
for various possible fast neutron spectra
and the resulting $\epsilon_1$ and $r$
for neutrons created in the lab walls were calculated. 
The same calculation was performed for neutrons 
created in the passive detector shielding by untagged muons;
in this case, the expected yield is 
much smaller, being only a few percent of that from the walls.
The simulated spectra of spallation neutrons are 
chosen to span the wide range of predictions
quoted in literature.

A value for $\epsilon_1$ of $1.14\pm0.07$ is found
after averaging over spectra,
implying that the spallation proton recoil background is 
essentially symmetric like the other backgrounds. 
Upon simulating the possible spectra, 
the quantity $(1-\epsilon_1)r=0.1\pm0.05$ is found to vary little.

The yield and spectrum of neutrons from muon capture
are reasonably well understood. Since 
these neutrons tend to be lower in energy,
only those created in the vicinity of the detector have any efficiency for 
creating background. Knowing the veto inefficiency to miss 
stopping muons ($4 \pm 1$)\%, the capture rate in water surrounding the 
detector and its contribution to the background can be estimated
using Monte Carlo simulation.
Overall this proton recoil background appears to be symmetric as well, 
$\epsilon_1=0.77\pm0.32$, 
meaning that the subtraction also strongly rejects this 
background.
The uncertainty of the residual background $(1-\epsilon_1)B_{\rm pn}$
is conservatively estimated to be about 160\%, corresponding to 
$\sim$4\% error on $N_{\nuebar}$.

\begin{figure}[htb!!!]
\centerline{\epsfxsize=3.5in \epsfbox{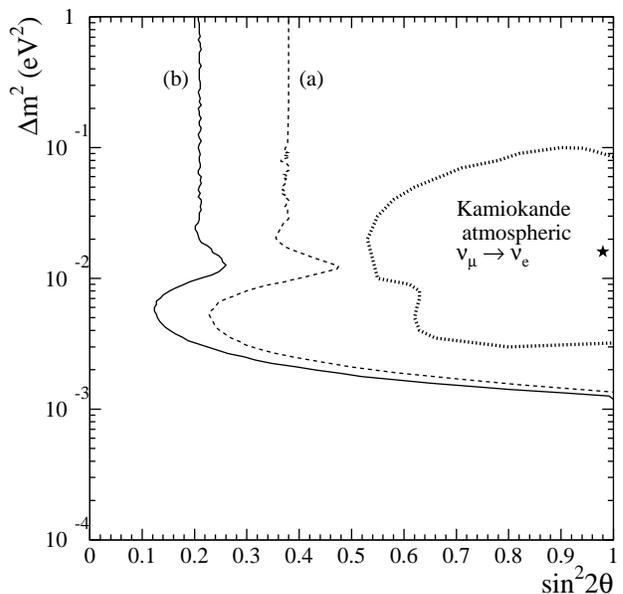}}
\caption{90\% CL limits of oscillation parameters
for (a) the {\em reactor power} analysis, which fits for background based 
on varying power levels, and (b) the {\em swap} analysis, 
which directly subtracts background from the data set. 
The Kamiokande $\nu_{\rm e}$--$\nu_\mu$ 
atmospheric neutrino 90\% CL (dashed line) and best fit (star)
are also shown.}
\label{fig:exclyf}
\end{figure}

The results of this analysis are summarized in Table~\ref{tab:swap}.
Overall 
\begin{equation}
\frac{R_{\rm obs}}{R_{\rm calc}}=1.04\pm 
0.03({\rm stat.})\pm0.08({\rm sys.}).
\end{equation}
The background estimates returned by the reactor power analysis
in Table~\ref{tab:and}
compare well with the results of the {\em swap}
analysis.
The 90\% CL region for this analysis follows the same $\chi^2$ 
formula, Eqn.~\ref{eq:chisqand}, as for the 
reactor power analysis but uses the background estimated by 
the {\em swap} method subtraction instead of minimizing the function 
with respect to background.
Again, we find no evidence for neutrino oscillations and the excluded 
region for this analysis is shown in Fig.~\ref{fig:exclyf}, curve (b).

\begin{table}
\begin{tabular}{|l|c|c|c|c|}
Analysis & On-Off & Spectrum & Power & Swap \\ 
\hline
$\chi^2_{\rm best}/{\rm n.d.f.}$   & 17.9/13 & 16.9/13 & 317.6/325           & 0.9/1 \\ \hline
$\alpha_{\rm best}$                &   1.00  &   1.99  &     1.02            & 1.03  \\ \hline
$\chi^2_{\rm no.osc}/{\rm n.d.f.}$ & 17.9/15 & 17.9/15 & 317.7/327           & 0.9/3 \\ 
\end{tabular}
\caption{The best fit to the data for the oscillation parameters
of the two flavor model for each of the analyses presented here:
{\em on-off, spectral shape, reactor power, and swap}.
The $\chi^2$ of the no-oscillation hypothesis
is also tabulated for each analysis.
For the  
{\em reactor power} analysis
all runs with less than 10 $\nuebar$ candidates were combined
with adjacent runs, leaving 327 d.o.f. See text above for an explanation of the ``Spectrum'' analysis.}
\label{tab:chisq}
\end{table}

\subsection{Systematic uncertainties}
The systematic uncertainties have three sources: the 
prediction of expected 
$\nuebar$ interactions, the efficiency estimate,
and, for the {\em swap} analysis, the $B_{\rm pn}$ estimate.
The expected $\nuebar$ uncertainty is dominated by 
the conversion of fission rates into neutrino fluxes, which 
relies on direct empirical measurements of $\beta$ spectra 
emitted by the isotopes. The Bugey experiment~\cite{Bugey_94}, which directly 
measured the neutrino flux and energy spectrum emitted by 
a reactor at short baseline, found agreement within 3\% using 
the same methods; the 3\% value is used 
here as the estimated uncertainty.

The efficiency uncertainty can be further subdivided into 
that arising from direct comparisons of Monte Carlo e$^+$ and 
neutron efficiency from calibration measurements and that 
arising from the selection cuts themselves. 
The calibration runs taken with the positron and neutron sources,
when compared with Monte Carlo simulations, shows overall 
agreement across all locations of better than 1\% in the efficiency 
predictions. However, the run-by-run agreement 
was at a level of 4\% for neutrons and 11\% for positrons.
Since the $^{22}$Na source is similar to the 
inverse beta decay signal with the e$^+$ close to 
detector threshold, the positron efficiency 
uncertainty over 
the entire $\nuebar$ spectrum 
was estimated to be closer to 4\% in any 
particular location.
These run-by-run variations are then used as our systematic 
uncertainties in the $\nuebar$ efficiency.

To test the robustness of the event selection,
each cut is varied within a reasonable range and
variations of the ratio between data and Monte Carlo are examined.
In order to take into account correlations 
all cuts were varied simultaneously by
randomly sampling a multidimensional 
{\em cut space}.  The {\em rms} of the resulting 
ratio of observed/expected is given as the selection cut 
uncertainty. 

The {\em swap} method analysis has 
a somewhat smaller uncertainty for the selection cuts variation
as the subtraction tends to cancel out systematics.
However, the {\em swap} analysis uses a Monte Carlo estimate of 
the proton recoil background. Due to limited Monte-Carlo statistics and
the uncertainty in the fast-neutron energy spectrum, a 
4\% uncertainty is assigned to the neutrino signal. 
All of the systematic uncertainties are summarized in 
Table~\ref{tab:sys}.
The total systematic uncertainty is obtained by adding the individual 
errors in quadrature.
\begin{table}[htb!!!]
\begin{tabular}{|l|c|c|}
Error Source &  On Minus Off(\%) & Swap(\%) \\ \hline
e$^+$ efficiency                  &       4       &       4       \\
n  efficiency                     &       3       &       3       \\
$\bar\nu_{\rm e}$ flux prediction &       3       &       3       \\
$\bar\nu_{\rm e}$ selection cuts  &       8       &       4       \\
$B_{\rm pn}$ estimate             &      ---      &       4       \\
\hline
Total                             &      10       &       8       \\
\end{tabular}
\caption{Summary of the systematic uncertainties.}
\label{tab:sys}
\end{table}

The development of a second simulation and event reconstruction proved
to be helpful in understanding systematic uncertainties of 
the analyses due to the algorithms chosen.
For comparison the results 
for the {\em on-off} analysis from both reconstructions are
shown in Tables~\ref{tab:onoff} and~\ref{tab:al}.
An independent analysis of systematic errors was performed
for the second reconstruction, 
similar to the method 
described above, giving comparable results.

\section{Conclusion}

In conclusion, the data taken thus far from the Palo Verde experiment 
show no evidence for $\nuebar\rightarrow\bar\nu_x$ oscillations.
This result, along with the results reported by Chooz\cite{Apollonio:1999ae} 
and Super-Kamiokande\cite{Fukuda:1998mi}, excludes 
two family $\nu_\mu$--$\nu_{\rm e}$ mixing as being responsible for the 
atmospheric neutrino anomaly as originally reported by 
Kamiokande\cite{Fukuda:1994mc}.
Later results of Super-Kamiokande, in particular 
data on the
zenith angle distribution of muons and electrons,
suggest that muon neutrinos $\nu_{\mu}$ strongly mix
with either $\nu_{\tau}$ or with a fourth flavor of neutrino
sterile to weak interaction. 
Clearly it is becoming important to include at least 
three neutrino flavors when studying results from oscillations
experiments.
\begin{figure}[htb!!!]
\centerline{\epsfxsize=3.5in \epsfbox{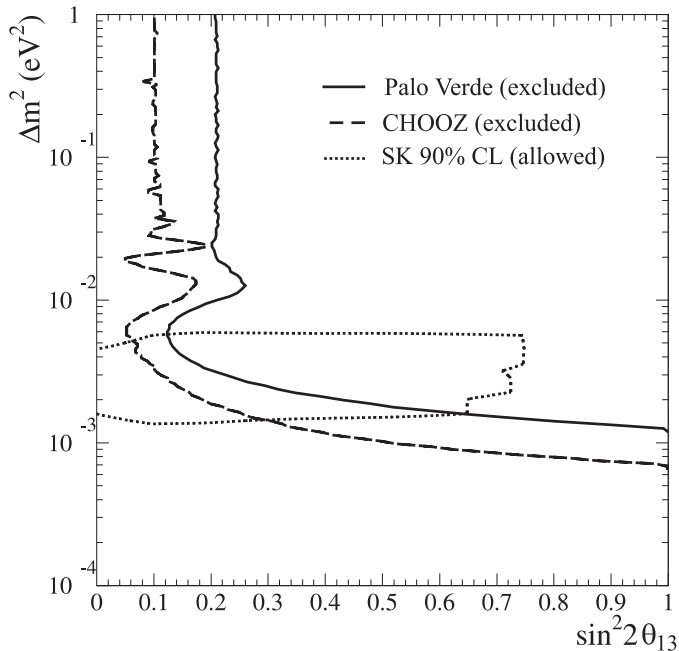}}
\caption{Exclusion plot showing the $allowed$ region of $\theta_{13}$
and $\Delta m^2$ based on the Super-Kamiokande preliminary analysis
(the region $inside$ the dotted curve). The region
$excluded$ by the neutrino reactor experiments 
are to the right of the corresponding dashed and continuous curves.}
\label{fig:SK}
\end{figure}

The most general approach would involve 
five unknown parameters, three
mixing angles and two independent mass differences. However,
an intermediate approach consists of a simple generalization
of the two flavor scenario, assuming that
$m_3^2 \gg m_1^2, m_2^2$ (i.e.
$\Delta m_{13}^2 = \Delta m_{23}^2 = \Delta m^2$, while 
$ \Delta m_{12}^2 \simeq 0$). In such a case the mixing angle
$\theta_{12}$ becomes irrelevant and one is left with
only three unknown quantities:
$ \Delta m^2, \theta_{13}, {\rm and} ~\theta_{23}$.
With this parameterization the $\bar{\nu}_e$ disappearance
is governed by
\begin{equation}
P(\bar{\nu}_e \rightarrow \bar{\nu}_x) = 
\sin^2 2\theta_{13} \sin^2 \frac{\Delta m^2 L}{4E_{\nu}} ~,
\end{equation}
while the $\nu_{\mu} \rightarrow \nu_{\tau}$ oscillations
in this scenario responsible for the atmospheric neutrino results,
are described by
\begin{equation}
P(\nu_{\mu} \rightarrow \nu_{\tau}) = 
\cos^4 \theta_{13} \sin^2 2\theta_{23}  
\sin^2 \frac{\Delta m^2 L}{4E_{\nu}} ~.
\end{equation} 

A preliminary analysis of the atmospheric neutrino data
based on these assumptions has been performed \cite{SK3} and its
results are shown in Fig. \ref{fig:SK} for the 
$\bar{\nu}_e$ disappearance channel. One can see that while
the relevant region of the mass difference $\Delta m^2$
is determined by the atmospheric neutrino data, the
mixing angle $\theta_{13}$ is not constrained very much. 
Here the reactor neutrino results play a decisive role.

We plan to continue taking data through Summer of 2000,
which will provide two additional reduced flux refueling
periods.

\section*{Acknowledgments}
We would like to thank the Arizona Public Service Company for
the generous hospitality provided at the Palo Verde plant.
The important contributions of M.~Chen, R.~Hertenberger, K.~Lou, and 
N.~Mascarenhas in the early stages of this project are gratefully 
acknowledged.
We thank K.~Scholberg for illuminating discussions on the Super-Kamiokande
three flavor analysis.
We are indebted to 
J. Ball, B.~Barish, R.~Canny, A.~Godber, 
J.~Hanson, D.~Michael, C.~Peck, C.~Roat, N.~Tolich, and A.~Vital 
for their help.
We also acknowledge the generous financial help from
the University of Alabama, Arizona State University, 
California Institute of Technology, and 
Stanford University. Finally, our gratitude goes to CERN, DESY, FNAL,
LANL, LLNL, SLAC, and TJNAF who at different times provided us with 
parts and equipment needed for the experiment.

This project was supported in part by the Department of Energy. 
One of us (J.K.) received support from the Hungarian OTKA fund,
and another (L.M.) from the ARCS Foundation.

\end{document}